\documentclass[usegraphicx,useAMS,usenatbib]{mn2e}
\usepackage{graphicx}
\usepackage{ulem}
\usepackage{amssymb}
\usepackage{multirow}
\usepackage{soul}

\def\ph{PHOX}
\def\tilde{~}
\def\fig{Fig.\,}
\def\eq{Eq.\,}
\def\sec{Section~}
\def\tab{Table~}

\def\mfive{M_{500}}
\def\rfive{R_{500}}

\def\xspec{{\tt XSPEC}}

\def\apec{{\tt APEC}}

\def\wabs{{\tt WABS}}

\def\msun{\rm{~M_{\odot}}}
\def\cm{\rm{~cm}}
\def\ms{\rm{~Ms}}
\def\ks{\rm{~ks}}
\def\ergs{\rm{~erg/s}}

\def\gpc{\rm{~Gpc}}
\def\mpc{\rm{~Mpc}}
\def\kpc{\rm{~kpc}}
\def\kev{\rm{~keV}}

\def\om{\Omega_{M,0}}
\def\oml{\Omega_{\Lambda,0}}
\def\omb{\Omega_b}

\title[X-ray MUSIC of Galaxy Clusters]{The MUSIC of Galaxy Clusters II:  
X-ray global properties and scaling relations}

\author[V. Biffi et al.]{V. Biffi$^{1,2}$\thanks{E--mail:
veronica.biffi@uam.es, biffi@sissa.it}, F. Sembolini$^{1,3}$, 
M. De\,Petris$^{3}$, R. Valdarnini$^{2,4}$, G. Yepes$^{1}$, S. Gottl\"ober$^{5}$\\
$^{1}$Departamento de F\'isica Teorica, Universidad Aut\'onoma de Madrid, Ciudad Universitaria de Cantoblanco, 28049 Madrid, Spain\\
$^{2}$SISSA - Scuola Internazionale Superiore di Studi Avanzati, Via Bonomea 265, 34136 Trieste, Italy\\
$^{3}$Dipartimento di Fisica, Sapienza Universit\'a di Roma, Piazzale Aldo Moro 5, I-00185 Roma, Italy\\
$^{4}$INFN, Trieste - Iniziativa Specifica QGSKY, Italy\\
$^{5}$Leibniz-Institut f\"ur Astrophysik, An der Sternwarte 16, 14482 Potsdam, Germany\\
}

\begin{document}

\pagerange{\pageref{firstpage}--\pageref{lastpage}} \pubyear{...}
\maketitle
\label{firstpage}
\begin{abstract}
We present the X-ray properties and scaling relations of a large
sample of clusters extracted from the Marenostrum MUltidark
SImulations of galaxy Clusters (MUSIC) dataset. 
We focus on a sub-sample of 179 clusters at redshift
$z\sim0.11$,  with $3.2\times 10^{14}h^{-1}\msun < M_{vir} < 2\times 10^{15}h^{-1}\msun$, complete in mass. We
employed the X-ray photon simulator PHOX to obtain synthetic Chandra
observations 
and derive observable-like global properties of the intracluster
medium (ICM), as X-ray temperature ($T_X$) and luminosity ($L_X$).  
$T_X$ is found to slightly under-estimate the true mass-weighted
temperature, although tracing fairly well the cluster total mass. 
We also study the effects of $T_X$ on scaling relations with cluster 
intrinsic properties: total ($\mfive$ and gas $M_{g,500}$ mass; integrated
Compton parameter ($Y_{SZ}$) of the Sunyaev-Zel'dovich (SZ) thermal
effect; $Y_X=M_{g,500}\,\,T_X$. We confirm that
$Y_X$ is a very good mass proxy, with a scatter on $\mfive-Y_X$ and
$Y_{SZ}-Y_X$ lower than $5\%$. 
The study of scaling relations among X-ray, intrinsic and SZ
properties indicates that simulated MUSIC clusters 
reasonably resemble the self-similar prediction, especially for
correlations involving $T_X$. The observational approach also allows for
a more direct comparison with real clusters, from which we
find deviations mainly due to the physical description of the ICM,
affecting $T_X$ and, particularly, $L_X$. 
\end{abstract}
\section{INTRODUCTION}\label{sec:intro}
In the last decades, X-ray observations of galaxy clusters have 
continuously provided us with precious information on their 
intrinsic properties and components.
In particular, the X-ray emission from the diffuse intracluster medium (ICM)
has been proved to be a good tracer of both the physics governing
the gaseous component and the characteristics of the underlying potential well,
basically dominated by the non-luminous, dark matter \cite[see][]{sarazin1986}.
\\
A reliable estimate of their total mass represents a fundamental goal of 
astrophysical and cosmological investigations, 
as the accurate weighing of clusters is also crucial to use them as 
cosmological probes, e.g. via number counts \cite[see, e.g.][and references therein]{allen2011}.
In the X-ray band clusters are very bright sources relatively easy to detect
out to high redshifts and constitute therefore a powerful tool to select large 
samples of objects for cosmology studies.
However, the mass determination via X-ray observations is mainly possible for 
well resolved, regular, nearby galaxy clusters, for which ICM density and temperature 
profiles are measurable with good precision and the Hydrostatic Equilibrium (HE) 
hypothesis can be safely applied.
\\
As widely discussed in the literature
\cite[][]{rasia2006,nagai2007,lau2009,meneghetti2010,suto2013},
the hydrostatic mass can mis-estimate the true total mass by
a factor up to $20-30$ percents, due to underlying
erroneous assumptions
(e.g. on the dynamical state of the system, 
on the ICM non-thermal pressure support,
on the models used to deproject observed density and temperature profiles,
or on cluster sphericity).
\\
In many cases, especially at high redshift or for more disturbed, irregular systems,
when the hydrostatic X-ray mass cannot be inferred reliably,
mass proxies are commonly employed to obtain indirect mass estimates.
Scaling relations between global cluster properties can be invoked to this purpose, 
offering a substitute approach to derive the total mass from other 
observables, e.g. obtained from the X-ray band or 
through the thermal Sunyaev Zel'dovich \cite[SZ;][]{sz1970,sz1972} effect.
\\
In fact, the simple scenario of the gravitational collapse,
by which gravity is dominating the cluster formation process,
predicts a self-similar scaling of basic cluster observables with their mass
\cite[][]{kaiser1986}.
Observations of galaxy clusters have confirmed the presence of correlations among 
cluster properties, and with mass, although indicating in some cases a certain level 
of deviation from the expected self-similar slopes.
The main reason for this deviation is that non-gravitational processes on
smaller scales (e.g. cooling, dynamical interactions,
feedback from active galactic nuclei -- AGN --)
do take place during the assembly of clusters and have
a non-negligible effect on their energy content.
In this respect, a remarkable example is represented by the X-ray 
luminosity--temperature relation,
which is commonly observed to be significantly steeper than expected
\cite[][]{white1997,markevitch1998,arnaud1999,ikebe2002,ettori2004_obs,maughan2007,zhang2008,pratt2009}.
\\
In order to employ scaling relations to infer masses, also the scatter
about the relations has to be carefully considered:
the tighter the correlation, the more precise can be the mass estimate.
Therefore, investigating the intrinsic scatter that possibly exists for
correlation with certain properties is extremely useful in order to individuate
the lowest-scatter mass proxy among many observable properties \cite[e.g.][]{ettori2012}.
This is the case, for instance, of the integrated Compton parameter, $Y_{SZ}$,
a measure of the thermal SZ signal which
has been confirmed to closely trace the cluster total mass 
by both simulations and observations
\cite[][]{dasilva2004,nagai2006,morandi2007,bonamente2008,comis2011,sembolini2012,kay2012,planckIIIi}.
The physical motivation for this is that $Y_{SZ}$ is related
to the ICM pressure (integrated along the line of sight), or equivalently to
its total thermal energy, and thus to the depth of the cluster potential well.
Likewise, another remarkably good candidate is also
the X-ray-analog of the $Y_{SZ}$ parameter, $Y_X$,
which was introduced by \cite[][]{kravtsov2006} and similarly
quantifies the ICM thermal energy by the product of gas mass and spectroscopic temperature.
Therefore, $Y_X$ correlates strictly with $Y_{SZ}$, but also with total mass,
given a fortunate anti-correlation of the residuals in temperature and gas mass.
\\
As widely explored in the literature, numerical hydrodynamical simulations
can be as precious as observations in unveiling the
effects of a number of physical processes on the global properties
and self-similar appearance of galaxy clusters 
\cite[see, e.g., recent reviews by][]{borgani2011,kravtsov2012}. 
Current hydrodynamical simulations can be further exploited 
when results are obtained in an observational fashion, 
which makes the results more directly
comparable to real data, e.g. from the X-ray band
\cite[e.g.][]{mathiesen2001,gardini2004,mazzotta2004,rasia2005,rasia2006,valdarnini2006,kravtsov2006,nagai2007,jeltema2008,biffi2012_1,biffi2013,biffiAN}
or weak lensing 
\cite[e.g.][]{meneghetti2010,rasia2012}.
Under this special condition, projection and instrumental effects, 
unavoidable in real observations, can be limited and explored 
for the ideal case of simulated clusters, as
the intrinsic properties can be calculated exactly from the simulation.
\\
In return, simulations themselves can take advantage of such technique,
as mis-matches between theoretical definitions of observable properties,
used in numerical studies, can be overcome 
\cite[see, e.g., studies on the ICM X-ray temperature by][]{mazzotta2004,valdarnini2006,nagai2007}
and the capability of the implemented physical descriptions
to match real clusters can be better constrained
\cite[e.g.][]{puchwein2008,fabjan2010,fabjan2011,biffi2013}.
\\
This is the approach we intend to follow in the present work.
Specifically, we study the
Marenostrum MUltidark SImulations of galaxy Clusters (MUSIC) data-set
of cluster re-simulations by means of synthetic X-ray observations obtained
with the virtual photon simulator PHOX \cite[][]{biffi2012_1}.
The scope of this analysis is to extend
the study carried on by
\cite{sembolini2012} to the X-ray features and scaling relations
of the MUSIC clusters, thereby providing a more complete picture of
this simulated set with respect to their baryonic properties.
In fact, X-ray observables are highly susceptible to the complexity of
the ICM physical state (e.g. to its multi-phase structure) and can be
more significantly affected by the numerical description of the
baryonic processes accounted for in the simulations (particularly
cooling, star formation and feedback mechanisms).
\\
The paper is organized as follows.
\\
In \sec\ref{sec:sims} we present the sub-sample of simulated objects considered
for the present study, whereas the generation and analysis of synthetic X-ray 
observations are described in \sec\ref{sec:x-obs}.
In \sec\ref{sec:results} we discuss the main results.
Specifically, X-ray observables are analysed and compared to theoretical estimates
derived directly from the simulations (\sec\ref{sub:xray_properties}).
In \sec\ref{sub:scal_rel} we present instead 
mass-observable correlations (\sec\ref{sec:mt}, \sec\ref{sec:ml}, \sec\ref{sec:yx_rel}),
as well as pure X-ray (\sec\ref{sec:lt_rel})
and mixed X-ray/SZ (\sec\ref{sec:x-sz}, \sec\ref{sec:ylt_rel}, \sec\ref{sec:yyx_rel}) 
scaling relations.
These are analysed and discussed with respect to
the effects of the observational-like approach, 
the X-ray temperature determination,
the compatibility with the expected self-similar scenario and 
the comparison against observational and previous numerical findings
(\sec\ref{sec:comparison}).
\\
Finally we draw our conclusions in \sec\ref{sec:discussion}.

\section{Simulated sample of galaxy clusters}\label{sec:sims}
The numerical simulations employed in the present work
are part of the MUSIC project, in particular of the MultiDark sub-set of
re-simulated galaxy clusters \cite[MUSIC-2, see][]{sembolini2012}.
\\
The MultiDark simulation (MD) is a dark-matter-only N-body simulation
performed with the adaptive refinement tree (ART) code \cite[][]{kravtsov1997},
resolved with $2048^3$ particles in a $(1 h^{-1}\gpc)^3$ volume
\cite[][]{prada2012}.
The cosmology assumed refers to the best-fit parameters obtained from the
WMAP7+BAO+SNI data, i.e. $\om = 0.27$, $\omb = 0.0469$, $\oml = 0.73$, 
$h_0=0.7$, $\sigma_8 = 0.82$ and $n=0.95$ \cite[][]{komatsu2011}.
\\
{\bf MUSIC-2 re-simulated clusters:} 
a complete, mass selected, volume limited sample of 282 clusters
has been extracted from a low resolution ($256^3$ particles)
run of the MD simulation.
Namely, it comprises all the systems in the $(1 h^{-1}\gpc)^3$ volume 
with virial mass larger than $10^{15}h^{-1}\msun$ at redshift $z=0$.
\\
Each one of the identified systems has been then re-simulated
with higher resolution and including hydrodynamics, 
within a radius of $6 h^{-1}\mpc$ from the
center of each object at $z=0$.
Initial conditions for the re-simulations were generated with the
zooming technique by \cite{klypin2001}.
The re-simulations were performed with the TreePM/SPH parallel code
GADGET \cite[][]{springel2001,springel2005}, including 
treatments for
cooling, star formation and feedback from SNe winds 
\cite[][]{springel2003}.
The final mass resolution for these re-simulations is
$m_{DM}=9.01\times 10^8 h^{-1}\msun$ and $m_{gas}=1.9\times 10^8
h^{-1}\msun$, respectively.
\\
This permitted to obtain a huge catalog of cluster-like haloes,
namely more than five hundreds objects more massive than $10^{14}
h^{-1}\msun$ at redshift zero.
\\
Snapshots corresponding to 15 different redshifts have been stored
between $z=9$ and $z=0$\footnote{The MUSIC-2 database is publicly available 
at {\tt http://music-data.ft.uam.es/}, as well as initial conditions.}.

\subsection{The simulated data-set}
The sub-sample of re-simulated galaxy clusters for which we want to
analyse X-ray properties is selected from the MUSIC-2 data-set
employed in the work presented by \cite{sembolini2012}.
\\
For the present analysis, we focus in particular on one snapshot
at low redshift, i.e. z=0.11.
This redshift corresponds to one of the first time records of the simulation output 
earlier than $z=0$\footnote{For the X-ray analysis with the photon simulator,
the snapshot $z=0$ is excluded for technical reasons, since
it would imply a formally null angular-diameter distance and, consequently,
divergent normalizations of the X-ray model spectra (i.e. infinitely large flux).}
and is suitable to investigate cluster properties in a relatively 
recent epoch with respect to the early stages of formation.
\\
Precisely, we select all the clusters matching the mass completeness
of the MUSIC-2 dataset at this redshift \cite[see][]{sembolini2012}, 
i.e. $M_{vir}(z=0.11) > 7.5 \times 10^{14}h^{-1}\msun$.
Additionally, we also enlarge the sub-sample in order to comprise
all the progenitors at $z=0.11$ of the systems with virial masses 
above the completeness mass limit at $z=0$ 
($M_{vir}(z=0) > 8.5 \times 10^{14}h^{-1}\msun$).
This practically extends the $z=0.11$ selection towards the 
intermediate-mass end.
\\
As a result, we obtain a volume-limited sample of 179 clusters
that is complete in mass at $z=0.11$, with 
$M_{vir}(z=0.11)$ spanning the range $[3.2-20]\times 10^{14}h^{-1}\msun$.
\section{The synthetic X-ray observations with PHOX}\label{sec:x-obs}
Synthetic X-ray observations of the galaxy clusters of the selected
sample have been performed by means of the X-ray photon simulator \ph{}
\cite[see][for an extensive presentation of the implemented approach]{biffi2012_1}.
\\
The cube of virtual photons associated to each cluster box has been
generated for the simulated snapshot at redshift $z=0.11$. 
\\
For each gas element in the simulation, X-ray emission has been
derived by calculating a theoretical spectral model with the
X-ray-analysis package \xspec{}\footnote{\scriptsize{See
    http://heasarc.gsfc.nasa.gov/xanadu/xspec/.}}
\cite[][]{xspec1996}.
In particular, we assumed the thermal \apec{} model \cite[][]{apec2001}, and also combined
this with an absorption model, \wabs{} \cite[][]{wabs1983}, in order
to mimic the suppression of low-energy photons due to Galactic
absorption. 
To this purpose, the equivalent hydrogen column density parameter has
been fixed to the fiducial value of $N_H = 10^{20}cm^{-2}$.
Temperature, total metallicity and density of each gas element, required to
calculate the spectral emission model, have been directly obtained
from the hydrodynamical simulation output.
\\
At this stage, fiducial, ideal values for collecting area and
observation time have been assumed, namely $A_{fid}=1000 \cm^2$ and
$\tau_{exp,fid}=1 \ms$.
\\
For the geometrical selection we have considered a cylinder-like region
enclosed by the $\rfive$ radius around each cluster.
The radius $\rfive$ is here defined as the radius encompassing a
region with an overdensity of $\Delta_{cr}=500$, with respect to the {\it critical} density of the
Universe, and in the following we will always use this definition
when referring to $\rfive$. 
This choice is motivated by our intention of comparing against the
majority of the observational and numerical works in literature,
which commonly adopt the same overdensity.
As for the projection, we consider a line of sight (l.o.s.) aligned with
the z-axis of the simulation box. 
\\
Finally, we assume a realistic exposure time of $100\ks$ and perform
synthetic observations for the ACIS-S detector of {\it Chandra}. This
is done with \ph{} Unit-3, by convolving the ideal photon lists extracted
from the selected regions with the ancillary response file (ARF) and
the redistribution matrix file (RMF) of the ACIS-S detector.
\\
Given the adopted cosmology, the $17' \times 17'$ field of view
(FoV) of Chandra corresponds at our redshift to a region 
in the sky of $2062.44 \kpc$ per side (physical units),
which would not comprise the whole $\rfive$ region for the majority of the
clusters in the sample.
Nevertheless, one can always assume to be ideally able to entirely
cover each cluster
with multiple-pointing observations and therefore we profit from the
simulation case to extract X-ray properties from within $\rfive$ for all the objects.

\subsection{X-ray analysis of Chandra synthetic spectra}\label{sec:xray_analysis}
The synthetic Chandra spectra generated with \ph{} have been re-grouped
requiring a minimum of 20 counts per energy bin. 
Spectral fits of the synthetic Chandra spectra, corresponding to
the $\rfive$ region of each cluster, have been performed over the
$0.5-10\kev$ energy band\footnote{for some clusters the band was
  actually restricted to a smaller energy band, depending on the
  quality of the spectrum} 
by using \xspec{} and adopting an absorbed (\wabs{}), thermally-broadened, \apec{} model, 
which takes into account a single-temperature plasma to model the ICM
emission.
In the fit, parameters for galactic absorption and redshift have been
fixed to the original values assumed to produce the observations,
while the other parameters were allowed to vary.
\\
For all the clusters in the sample, the best-fit spectral model
generally indicates a very low value of the total metallicity of the
plasma. This result is simply reflecting the treatment of the star
formation and metal production in the original input simulations,
which does not follow proper stellar evolution and injection of metal 
yields according to proper stellar life-times.
\\
We note that, even though the ICM in the clusters is most likely constituted by a
multi-phase plasma, the single-temperature fit results overall in reasonable
estimations, for all the clusters in our sample, as confirmed by the
$\chi^2$ statistics 
(see \fig\ref{fig:chi2_fits} in Appendix~\ref{appA}).
From the distribution of the reduced-$\chi^2$ values, we have in
fact $\chi^2 < 1.2$ for $\sim 70\%$ of the clusters in the sample.
%
\section{Results}\label{sec:results}
In this Section we present the X-ray properties and scaling relations of the 179 galaxy
clusters in the sample, at $z=0.11$, obtained from Chandra synthetic observations.
\subsection{X-Ray properties of the massive haloes}\label{sub:xray_properties}
Two interesting, global X-ray properties that we can directly extract from
spectral analysis are the luminosity and temperature of the ICM within
the projected $\rfive$. 
\subsubsection{The X-ray luminosity}\label{sub:lum}
\begin{figure}
\centering
\includegraphics[width=0.45\textwidth]{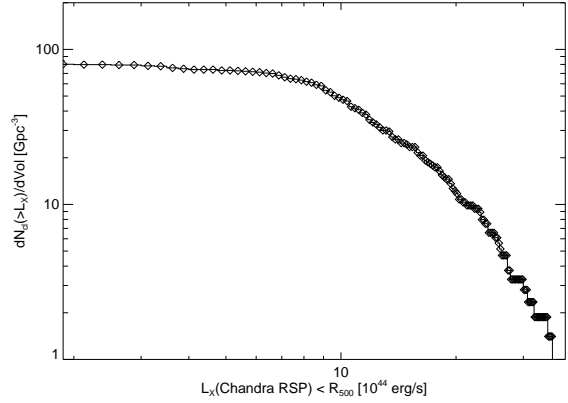}
\caption{Cumulative luminosity function for the sample of clusters. } 
\label{fig:lum_fnc}
\end{figure}

From the theoretical best-fit model to the synthetic data, we
calculate X-ray luminosities for the sample clusters in the entire
$0.5-10\kev$ band, 
as well as in the Soft and Hard X-ray bands,
i.e. $0.5-2\kev$ and $2-10\kev$, respectively (rest frame energies).
\\
Furthermore, we extrapolate the total X-ray luminosity to the maximum
energy band defined by the Chandra response matrix, i.e. $0.26-12\kev$ 
($0.28-13.3\kev$ rest frame).
Hereafter, we will refer to this quantity as the 
``bolometric'' X-ray luminosity.
\\
In \fig\ref{fig:lum_fnc} we show the cumulative luminosity function 
built from the ``bolometric'' X-ray luminosity within $\rfive$ of all the clusters in the sample,
${\rm d}{N}(>L_X)/{\rm d}{Vol}$ (for a volume corresponding to the simulation
box volume) 

\subsubsection{The X-ray temperature}\label{sec:temp}
From the analysis of the Chandra synthetic spectra,
we also measure the projected mean temperature within $\rfive$.
All the clusters of the sample have temperatures $T_X>2\kev$.
This temperature, usually referred to as ``spectroscopic'' temperature
can be compared to the temperature estimated from the simulation as
\begin{equation}\label{eq:t_sim}
 T_w = \frac{\Sigma_i w_iT_i }{\Sigma_i w_i},
\end{equation}
where the sums are performed over 
the SPH particles in the considered 
region of the simulated cluster.
The temperature associated to the single gas particle ($T_i$) 
is computed taking into account the multi-phase gas description 
following the model by \cite{springel2003}.
\\
The weight $w$ in \eq\ref{eq:t_sim} changes according to different theoretical definitions.\\
As commonly done, we consider: 
\begin{itemize}
 \item the {\bf mass-weighted} temperature, $T_{mw}$, where $w_i = m_i$;
 \item the {\bf emission-weighted} temperature, $T_{ew}$, where the emission is
$\sim \rho^2\Lambda(T)\sim \rho^2\sqrt{T}$ 
(the cooling function can be approximated by $\Lambda(T)\sim \sqrt{T}$ 
for dominating thermal bremsstrahlung) 
and therefore $w_i = m_i \rho_i \sqrt{T_i}$;
 \item the {\bf spectroscopic-like} temperature, $T_{sl}$, where $w_i = m_i \rho_i T_i^{-3/4}$ 
\cite[which was proposed by][as a good approximation of the spectroscopic temperature for systems with $T\gtrsim 2-3\kev$]{mazzotta2004}.
\end{itemize}
While computing the emission-weighted and spectroscopic-like temperatures,
we apply corrections to the particle density that account for the multi-phase gas model adopted
(consistently to what is done while generating the X-ray synthetic emission)
and we exclude cold gas particles, precisely those with temperatures $< 0.5\kev$.
\\
Among the aforementioned theoretical estimates, 
the mass-weighted temperature is the value that
most-closely relates to the mass of the cluster, 
directly reflecting the potential well of the system.
\\
\begin{figure}
\includegraphics[width=0.46\textwidth]{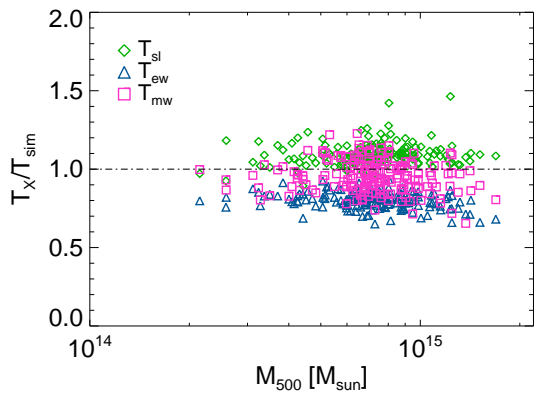}
\caption{Relations between the X-ray-derived temperature, obtained from
  spectral fitting of the synthetic spectra in the (0.5-10) keV
  band, and the different theoretical definitions of temperature
  estimated directly from the simulation: 
  $T_{mw}$ (mass-weighted, magenta squares), 
  $T_{ew}$ (emission-weighted, blue triangles), $T_{sl}$ (spectroscopic-like, green diamonds).
  Temperature ratios are 
  plotted as function of the cluster mass within $\rfive$ ($\mfive$).} 
\label{fig:temp}
\end{figure}
\begin{figure}
\includegraphics[width=0.46\textwidth]{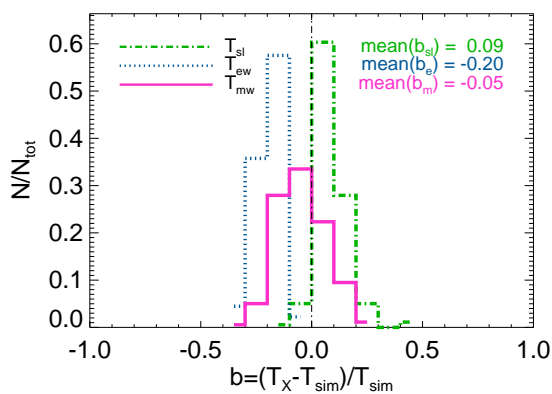}
\caption{Distribution of the deviations of the X-ray temperature from the
  simulation estimates, according to the three theoretical
  definitions: 
  $T_{mw}$ (solid), $T_{ew}$ (dotted), $T_{sl}$ (dash-dotted). 
  Colors as in \fig\ref{fig:temp}. The distributions are
  normalized to the total number of clusters in the sample.
  Mean value and standard deviation of each distribution are: 
  $(0.09\pm 0.07)$ for $T_{sim}\equiv T_{sl}$;
  $(-0.20\pm 0.05)$ for $T_{sim}\equiv T_{ew}$;
  $(-0.05\pm 0.10)$ for $T_{sim}\equiv T_{mw}$.} 
\label{fig:bias}
\end{figure}
%
Nevertheless, the various other ways of weighting the temperature for the
gas emission (such as $T_{ew}$ or $T_{sl}$) have been introduced in order to 
better explore the X-ray, observable properties of
simulated galaxy clusters and to ultimately compare against real observations.
Differences among these definitions and their capabilities to match the
observed X-ray temperature have been widely discussed in the literature 
\cite[e.g. ][]{mathiesen2001,mazzotta2004,rasia2005,valdarnini2006,nagai2007}.
\\
From the comparison shown in \fig\ref{fig:temp}, we also remark the differences
existing among the theoretical estimates and the expected spectroscopic 
temperature $T_X$, derived from proper spectral fitting.
The spectroscopic temperature refers, in this case, to the region
within the {\it projected} $\rfive$,
which
might introduce deviations due to substructures lying along the l.o.s..
Despite this, we expect it to be fairly consistent to the
global, 3D value, given the large region considered.
\\
In \fig\ref{fig:temp}, the ratio $T_X/T_{sim}$ is presented as a function of the true cluster
mass within $\rfive$, $\mfive$.
\\
Comparing to the 1:1 relation (black, dot-dashed line in the Figure), 
we note that there are discrepancies among the
values.
Overall, we conclude from this comparison that $T_X$ tends to
be generally higher than the value of $T_{sl}$.
Also, in perfect agreement to the findings of previous works
\cite[e.g.][]{mazzotta2004}, the spectroscopic temperature $T_X$ is on
average lower than the
emission-weighted estimate.
\\
It's interesting to notice, that in our case 
the discrepancy between $T_X$ and the true, 
dynamical temperature of the clusters, $T_{mw}$, 
is smaller than the deviation from either
emission-weighted or spectroscopic-like temperatures.\\
Nonetheless, with respect to the mass-weighted value,
we find that $T_X$ tends to be slightly biased low.
On one hand, this under-estimation of the true temperature by
the X-ray-derived measurement might be ameliorated via a further exclusion 
{\it a posteriori} of
cold, gaseous substructures in the ICM. 
On the other hand, a complexity in the
thermal structure of the ICM can persist 
(for instance, a broad temperature distribution,
or a significant difference in the temperatures of the two most prominent gas phases)
and eventually affect the resulting X-ray temperature, 
especially when a single temperature component is fitted to the integrated spectrum.
\\
As a test, we checked for the dependency of the bias on the $\chi^2_{red}$ of the fit
and found that there is a mild correlation (see \fig\ref{fig:ratio_chired} in Appendix~\ref{appA}).
Precisely, to higher $\chi^2_{red}$ values of the one-temperature spectral fit,
correspond on average lower $T_X/T_{mw}$ ratios (and similarly for $T_X/T_{ew}$).
Comparing $T_X$ to $T_{sl}$ this effect basically disappears, confirming that
they are almost equally sensitive to the complexity of the ICM thermal distribution.
\\
The observed under-estimation of the true temperature by $T_X$ 
is in agreement with findings from, e.g., early studies using
mock X-ray observations of simulated clusters by \cite{mathiesen2001},
but there is instead some tension with respect to other numerical studies
\cite[e.g.][]{nagai2007,piffaretti2008}. Nevertheless, as reported also by
\cite{kay2012}, who found results consistent with what observed in MUSIC clusters, 
the discrepancy might be due to the additional exclusion of resolved cold 
clumps in the X-ray analysis.
\\
In fact, for the set of MUSIC clusters, we observe that
a two-temperature model would generally improve the quality of the fit
(especially for the objects where the single-temperature fit provides $\chi^2_{red} > 1.2$),
better capturing the local multi-phase nature of the gas. 
However, the best-fit hotter component tends to {\it over}-estimate 
the mass-weighted temperature, introducing even in this case a significant 
bias in the results.
Moreover, this increases the overall scatter, particularly for colder, low-mass systems,
where it is more difficult to distinguish between the two temperature components.
Therefore, we decide to consider throughout the following analysis the results from the 
single-temperature best-fit models.
\\
Another, more quantitative, way of comparing the deviations of  
$T_{sl}$, $T_{ew}$ and $T_{mw}$ 
from $T_X$
is by confronting the distributions of the bias,
defined as:
\begin{equation}\label{eq:bias}
b = \frac{T_X - T_{sim}}{T_{sim}},
\end{equation}
shown in \fig\ref{fig:bias}. 
\\
From this we clearly observe that the distribution of $b$ for 
$T_{sim}\equiv T_{mw}$ shows the best agreement with the 1:1 relation,
although it is not symmetrical and rather biased toward negative deviations.
This corresponds to an average under-estimation by $T_X$ of $\sim 5$ percent,
over the all sample.
More specifically, we find that for almost $\sim 67\%$ of the clusters considered
$T_X$ under-estimates the true temperature of the system. 
\\
Emission-weighted and spectroscopic-like temperatures suggest instead 
more extreme differences and narrower distributions. 
$T_{ew}$ indicates a more significant mis-match with the X-ray value,
which tends to be smaller by a factor of $\sim 20\%$, with little dispersion.
$T_{sl}$ is instead closer to the X-ray temperature, although the 
distribution of the deviations is slightly biased to positive values of $b$,
indicating a typical over-estimation by $T_X$ of a few percents.
The mean value of each bias distribution is reported in the legend of \fig\ref{fig:bias}.
%
\begin{figure*}
\centering 
\includegraphics[width=0.49\textwidth]{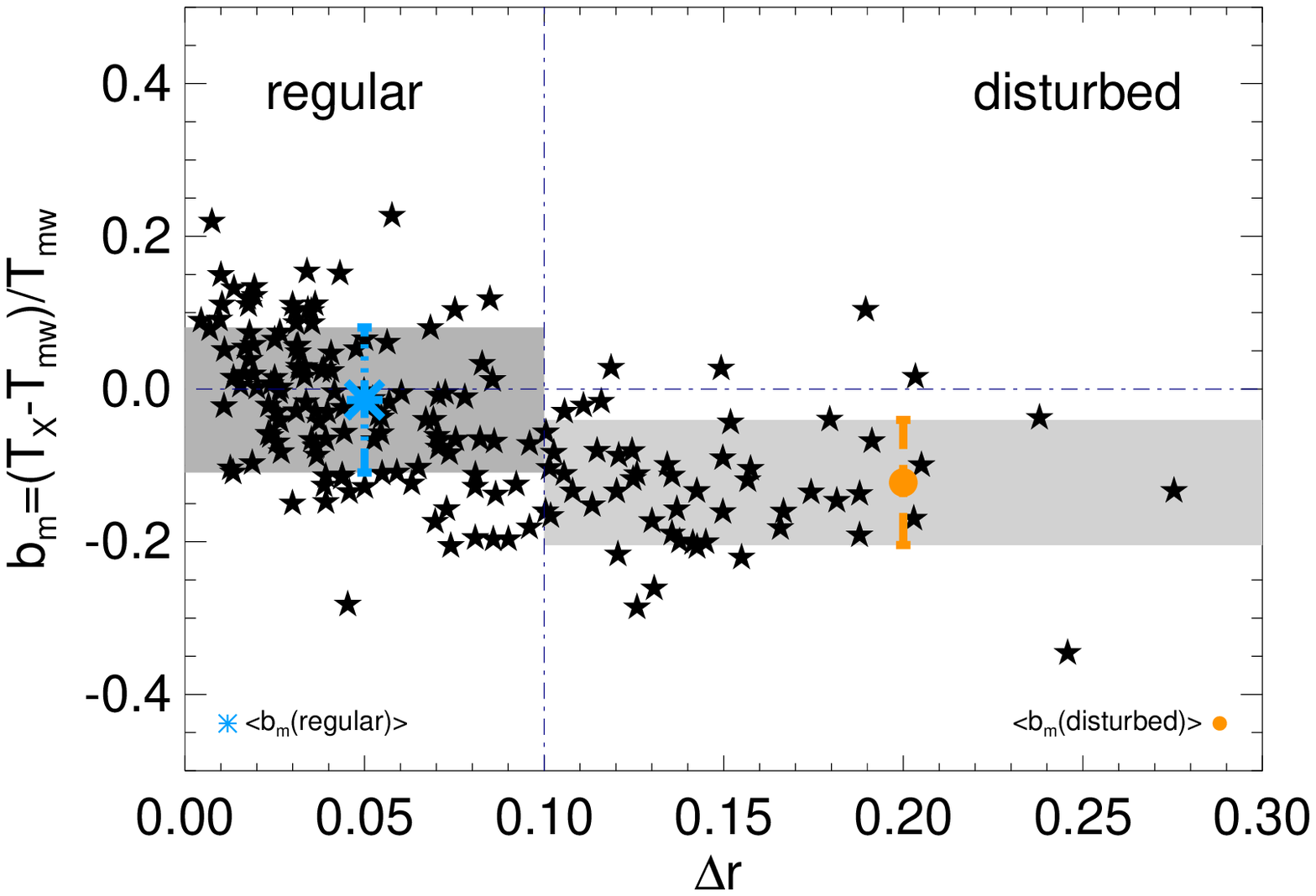}
\includegraphics[width=0.48\textwidth]{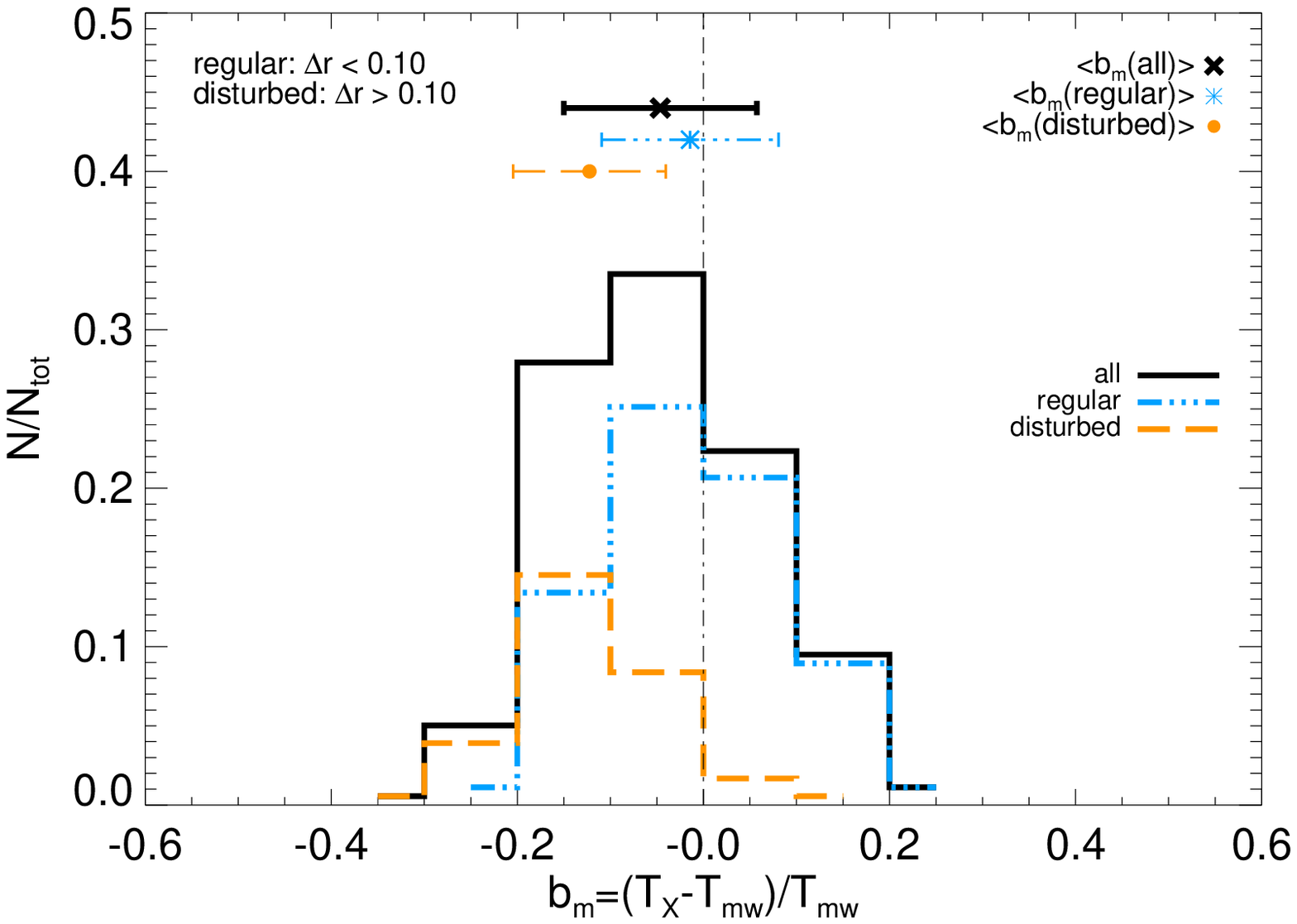}
\caption{{\it Left panel:} Dependence of the temperature bias ($b_m=(T_X -
    T_{mw})/(T_{mw})$) on the dynamical state of the cluster,
    quantified by the center-of-mass off-set ($\Delta r$).
    The dot-dashed vertical line marks the threshold chosen to separate
    regular clusters from disturbed ones.
    In each region of the plot the mean value of the bias $b_m$ and
    standard deviation are marked with a symbol (blue asterisk for regular clusters, orange filled circle
    for disturbed ones) and a shaded area.
    {\it Right panel:} The distribution of the bias $b_m$ is reported
    for the complete sample (black, solid line) and the regular (blue, dot-dot-dot-dashed
    line) and disturbed  (orange, dashed line) sub-sets. Also, the mean values, with 1-$\sigma$ errors, are
    reported for the three cases (black x, blue asterisk, orange filled circle,
   respectively).
}
\label{fig:bias_dr}
\end{figure*}
\subsubsection{Dependence of the temperature bias on the dynamical
  state}\label{sub:tbias_dr}
In order to investigate further the bias in the temperature
estimation, we concentrate particularly on the bias between
spectroscopic and mass-weighted values, i.e.
$b_m = (T_X - T_{mw})/(T_{mw})$ (see \eq\ref{eq:bias}),
and explore its relation to the global, intrinsic state of the cluster.
To this scope, we calculate from the simulations the center-of-mass
off-set, defined as
\begin{equation}
\Delta r = \frac{\|r_\delta - r_{cm}\|}{R_{vir}},
\end{equation}
namely the spatial separation between the maximum density peak
($r_\delta$) and the center of mass ($r_{cm}$), normalized to the
cluster virial radius ($R_{vir}$).
The choice to adopt this value to discriminate between {\it regular} and
{\it disturbed} clusters is related to the search for a quantity able to
describe the intrinsic state of the cluster, taking advantage of the
full three-dimensional information available in simulations.
The threshold adopted to divide the clusters into two sub-samples is
the fiducial value of $\Delta r_{th} = 0.1$ \cite[see, e.g.,][]{donghia2007}. 
This represents an upper limit
in the range of limit values explored in the literature
\cite[][]{maccio2007,neto2007,donghia2007,knebe2008}.
In our case, given the presence of the baryonic component, we decide
in fact to allow for a less stringent criterion \cite[see also][]{sembolini2013arxiv}.
\\
In \fig\ref{fig:bias_dr} we show the results of this test. 
The relation between observed bias and $\Delta r$ is shown in the left
panel of the Figure. The vertical dashed line marks the separation
threshold between regular and disturbed clusters.
We note that there is 
indeed a dependence of the $T_X-T_{mw}$ bias on the
dynamical state of the cluster,
with a general tendency for $b_m$ to increase with
increasing level of disturbance, quantified by the center-of-mass
off-set. 
More specifically, it is more
negative for higher values of $\Delta r$.
The filled circle and asterisk, and shaded areas, corresponding to the mean values
and standard deviations for the two subsamples, show indeed that the
bias distribution is centered very close to zero for the regular
clusters, while a more significant off-set is evident for the
disturbed sub-sample. 
The bias distributions for the two subsamples are shown more clearly in the
right panel of \fig\ref{fig:bias_dr} and compared to the global
distribution.
We note that, the bias calculated for the entire sample is basically
dominated by the regular clusters, which constitute the majority of
the haloes, 
given the threshold value adopted for $\Delta r$.
In particular, we find that for regular clusters $T_X$ approximates
to a few percents, on average, the true temperature of the cluster:
$<b_M(regular)>= -0.01\pm 0.01$.
The disturbed clusters, instead, are characterised by
$<b_M(disturbed)>= -0.12\pm 0.01$, indicating a stronger under-estimation. 
\\
Despite the difference in the mean values, we remark that the
distributions of the two populations are quite broad with respect to
the bias, as quantified by the standard deviations and shown also in
\fig\ref{fig:bias_dr} ($\sigma_{b_m}(regular)=0.10$ and $\sigma_{b_m}(disturbed)=0.08$).
\subsection{Global scaling relations}\label{sub:scal_rel}
\begin{figure*}
\centering
\includegraphics[width=0.49\textwidth]{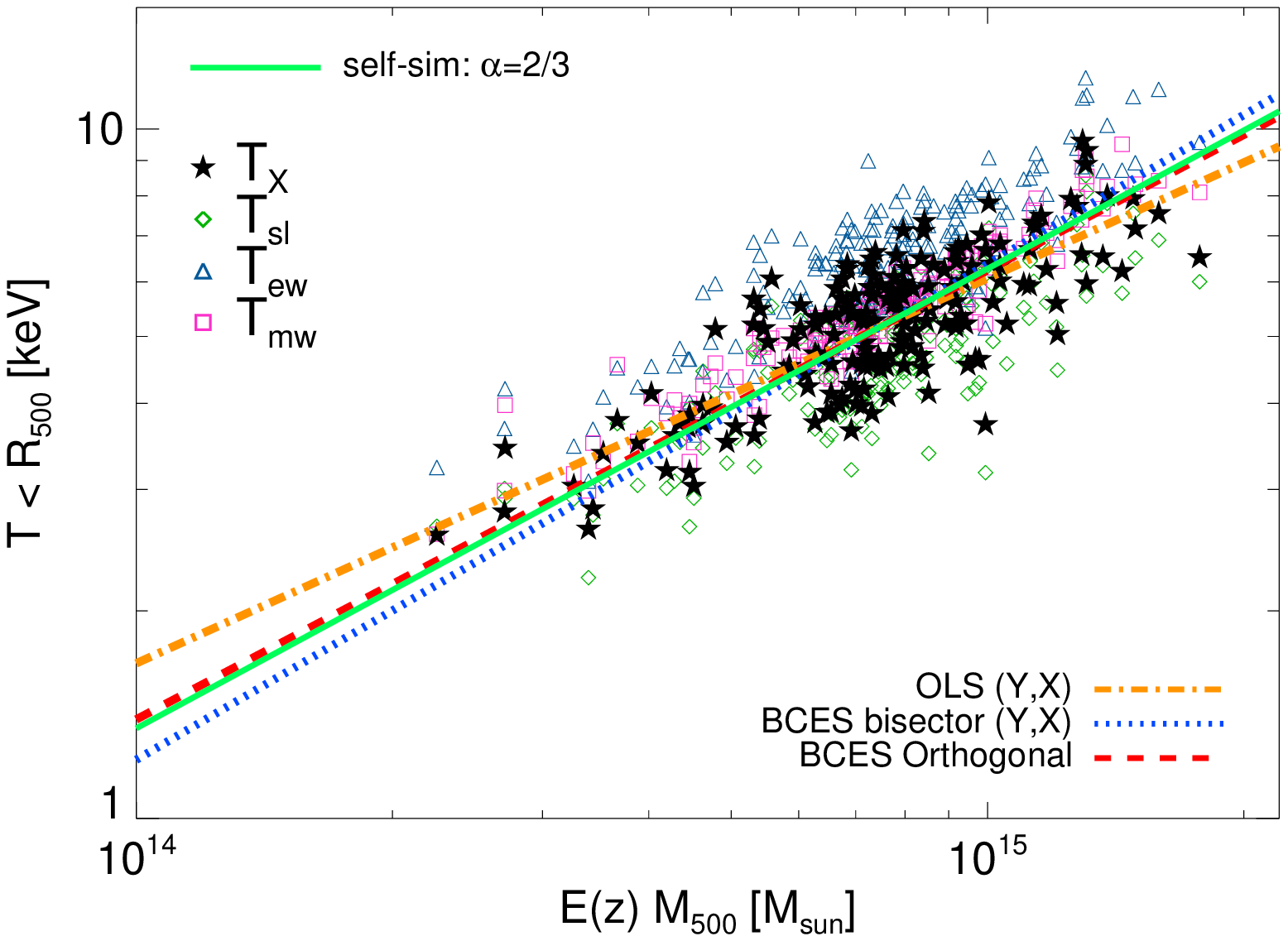}
\includegraphics[width=0.49\textwidth]{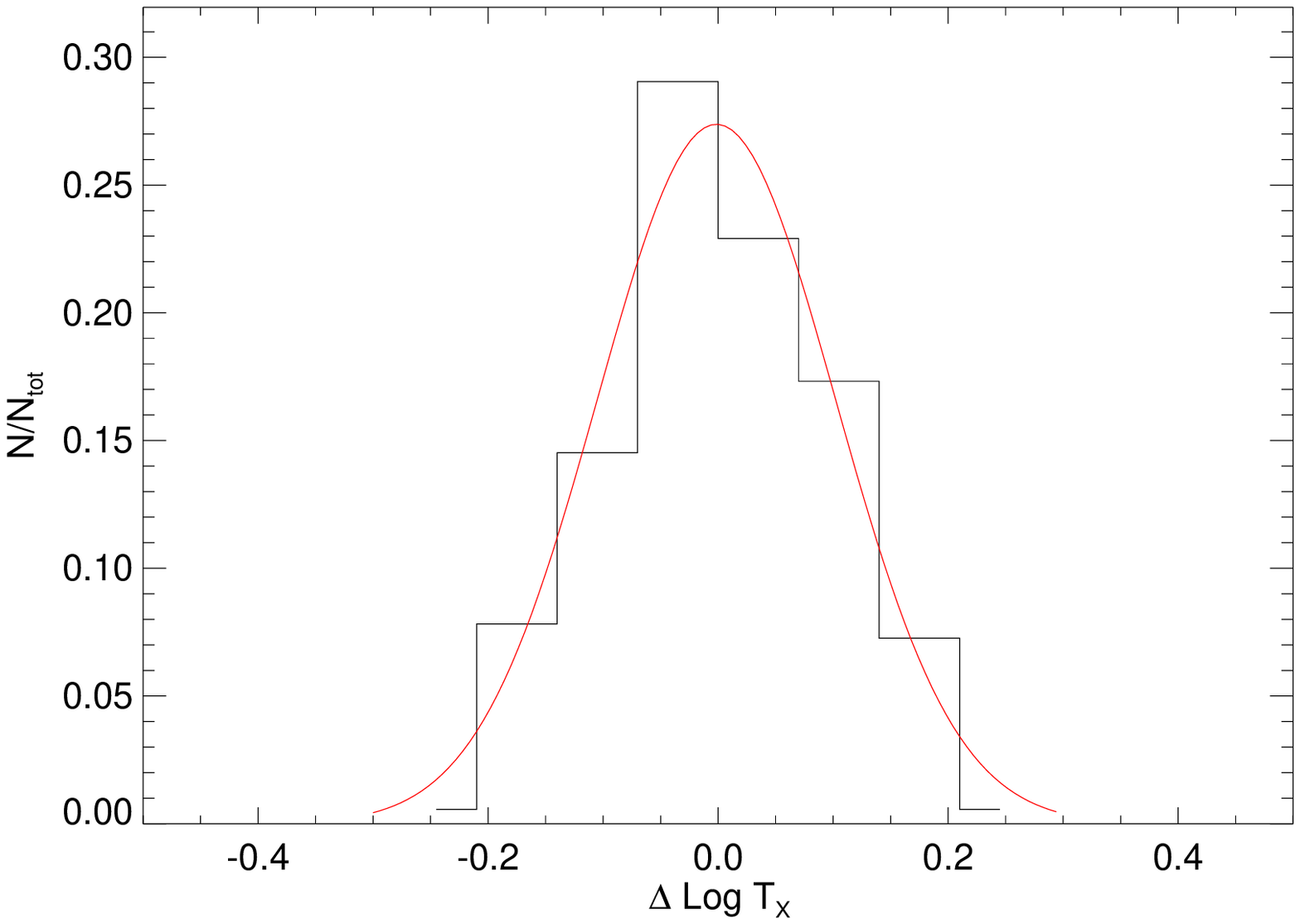}
\caption{{\it Left:} Temperature-mass relation. Open symbols refer to the
  three standard definitions of temperature used in simulations:
  $T_{mw}$ (magenta squares), $T_{ew}$ (blue triangles), $T_{sl}$ (green diamonds). 
  Black stars refer to the X-ray temperature, $T_X$, extracted from the synthetic
  spectra in the (0.5-10)~keV band. Over-plotted in orange the (OLS) best-fit
  relation for the $T_X - \mfive$ relation. For comparison, 
  we also report the Bisector and Orthogonal best-fit curves (as in the legend) 
  and the self-similar line, normalized to the data at $7.5\times 10^{14}\msun$, in mass.
  {\it Right:} distribution of the 
  residuals in the $T_X -\mfive$ relation (in ${\rm Log}T_X$).}
\label{fig:tm_rel}
\end{figure*}
In this Section we focus on cluster global scaling relations. 
We consider correlations among
X-ray quantities measured from the synthetic \ph{} observations (e.g. $L_X$, $T_X$),
properties estimated from the thermal SZ signal and
intrinsic quantities obtained from the numerical, hydrodynamical simulation 
data directly.
\\
Also, we aim to compare our findings with both current observational results
and theoretical expectations from the gravity-dominated scenario of cluster self-similarity.
\\
In this simplified model, the gravitational collapse giving birth to clusters of galaxies
determines entirely the global scaling of the system observable properties.
Precisely, the gas is assumed to be heated by the gravitational process only,
therefore depending uniquely on the scale set by the system total mass
(i.e. by the depth of its potential well), and on the redshift
\cite[see][]{kaiser1986}.
Under these assumptions, power law correlations for each set
of observables (Y,X) are expected, namely
\begin{equation}\label{eq:scal_rel}
{\rm Y} = C\,({\rm X})^\alpha,
\end{equation}
where $C$ and $\alpha$ are the normalization and the slope of the relation, respectively.
Throughout the following, we fit the data with linear relations in the
Log-Log plane\footnote{According to
  our notation, ${\rm Log\equiv log}_{10}$} of the general form
\begin{equation}\label{eq:scal_rel_log}
{\rm Log(Y)} = B + \alpha{\rm Log(X)},
\end{equation}
with $B={\rm Log}C$.
\\
The slope and the normalization are recovered via a minimization of the residuals
to the best-fit curve (further details on the minimization method adopted will be provided
on a case-by-case basis, in the following sections).
\\
Under this formalism, we also calculate the scatter in the Y variable as
\begin{equation}\label{eq:scatter}
\sigma_{{\rm Log Y}} = \left[\frac{\Sigma_{i=1}^N
    [{\rm Log}({\rm Y_i}) -
    (B + \alpha{\rm Log}({\rm X_i})) ]^2}{N-2}\right]^{1/2},
\end{equation}
where $N$ is the number of data points (for our analysis this is $N=179$, 
i.e. the number of clusters in the sample).
\subsubsection{Relation between temperature and mass}\label{sec:mt}
Here we discuss the relation between temperature and total mass
for the subsample of the MUSIC-2 clusters analysed in this work.
This is displayed in the left panel of \fig\ref{fig:tm_rel}.
\\
The differences that appear while comparing the spectroscopic
temperature $T_X$ to $T_{mw}$, $T_{ew}$ and $T_{sl}$, basically carry
the imprints of the differences existing among the three theoretical
estimates of temperature calculated directly from the simulations
(see \fig\ref{fig:tm_rel}, left panel).
Although a shift in temperatures is evident for the different data-sets
in \fig\ref{fig:tm_rel} (left panel), we note that for none of them
the spread in temperature shows any strong dependence on mass.
\\
Regarding $T_X$,
we observe that the observational-like temperatures obtained with PHOX
appear to be slightly more dispersed than the theoretical values.
This might reflect some contamination due to substructures residing along the line of sight
and within the projected $\rfive$, as well as the effect of single-temperature spectral fitting.
Even though not major, an increase in scatter and in deviation from self-similarity
is also expected as an effect of a more observational-like analysis.
\\
Nevertheless, the overall good correlation between mass and X-ray temperature ensures that
the latter behaves as a good tracer of the mass of
our clusters, even up to $\rfive$.
This is particularly interesting as the temperature is derived from
the cluster X-ray emission, while the mass considered here is the true
mass calculated from the simulation.
\\
As a step further, we recover the best-fit relation between
the spectroscopic temperature $T_X$ and $\mfive$.
As in \eq\ref{eq:scal_rel_log},
we fit the data in the Log-Log plane using the functional form
\begin{equation}\label{eq:tm_fit}
{\rm Log} (T_X) = {\rm Log}(C) + \alpha{\rm Log} (E(z)\,\mfive),
\end{equation}
in order to recover slope and normalization of the $T_X-\mfive$ scaling law.
\\
In \eq\ref{eq:tm_fit} (and hereafter), the function $E(z)^2=\om (1+z)^3 + \Omega_{\Lambda}$ 
accounts for the redshift and the cosmology assumed.
\\
In this case, we proceed with a simple ordinary least squares (OLS)
minimization method to calculate the slope and normalization of the
relation, since the mass is here the intrinsic, true value
obtained from the simulation and therefore it should be treated as
the ``independent'' variable.
As a result, we find a shallower slope ($\alpha = 0.56\pm 0.03$) than
predicted by the self-similar model 
($\alpha_{self-sim}=2/3$).
On the one hand, this shallower dependence might reflect the tendency of $T_X$ 
to under-estimate the true temperature of the system.
This is consistent, e.g., with results from numerical studies by \cite{jeltema2008}.
On the other hand, the minimization method itself could induce differences
in the results, especially when some intrinsic scatter in the relation is present.
Indeed, we find a steeper slope when the residuals on both variables are minimized, 
e.g. via the
bisector -- $\alpha = 0.72\pm 0.03$ --
or orthogonal -- $\alpha = 0.65\pm 0.04$ --
approaches
(Bivariate Correlated Errors and intrinsic Scatter -- BCES)
\footnote{We note here that we do not consider errors for the variables
  directly derived from the simulation (i.e. total mass, gas
  mass, $Y_{SZ}$, but also for $Y_X$), while they are accounted for in the
  relation between observational quantities, namely $T_X$ and $L_X$. 
  Nevertheless, we remark that the main important difference with respect to the OLS
  method is that the bisector or orthogonal approaches, the
  minimization accounts for the
  residuals in {\it both} X and Y variables,
  therefore providing potentially different best-fit slopes.}. 
For the purpose of comparison, we report in \fig\ref{fig:tm_rel} (left panel)
the best-fit curves for all the three methods, as well as the
self-similar relation (normalized to the data at $7.5\times 10^{14}~h^{-1}\msun$, 
in mass).
\\
For this sample, the scatter in ${\rm Log}T_X$
with respect to the best-fit relation is some percents, namely 
$\sigma_{{\rm Log}T_X}\sim 0.07$, calculated according to \eq\ref{eq:scatter}.
\\
In the right-hand-side panel of \fig\ref{fig:tm_rel} we show the
distribution of the OLS residuals for the $T_X - \mfive$ relation, in
${\rm Log}T_X$. This can be fitted by a Gaussian function,
centered on zero and with standard deviation $\sigma \sim 0.10$.
%
%
%
\subsubsection{Relation between luminosity and mass}\label{sec:ml}
\begin{figure}
\centering
\includegraphics[width=0.46\textwidth]{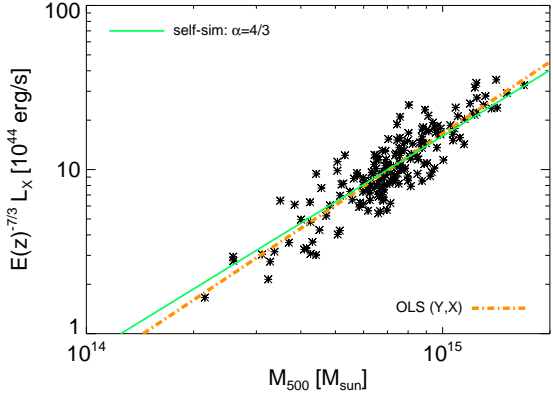}
\caption{Luminosity--mass relation for the total X-ray band
  covered by the Chandra detector response, to which we refer as
  ``bolometric'' X-ray luminosity. The over-plotted dot-dashed, 
  orange curve is the (OLS) best-fit relation to the data,
  while the solid green line marks the self-similar relation, 
  normalized to the data at $7.5\times 10^{14}~h^{-1}\msun$, in mass.} 
\label{fig:ml}
\end{figure}
As well as the temperature, also the X-ray luminosity ($L_X$) 
is expected to scale with the cluster mass 
\cite[see, e.g.][for a recent review on cluster scaling relations]{giodini2013}.
Therefore, we show here the $L_X-\mfive$ relation for our clusters,
within $\rfive$.
In the case of luminosity, we expect the signature of gas physics to play a major 
role, introducing both a deviation from self-similarity and a larger scatter.
Indeed, the X-ray emission of the ICM is much more sensitive to its thermal state,
e.g. to the multi-temperature components of the gas and to substructures.
Furthermore, the implementation of the complex processes governing the gas
physics, such as cooling, metal enrichment and feedback mechanisms, can certainly have a
non-negligible effect.
\\
Indeed, we observe a steeper correlation than expected and a larger scatter with respect 
to the temperature--mass relation, $\sigma_{{\rm Log}L_X}\sim 0.11$ (\sec\ref{sec:mt}).
The best-fit slope obtained from the OLS minimization method is
$\alpha \sim 1.45\pm 0.05$, with self-similarity predicting $\alpha_{self-sim}=4/3$.
Even though in the same direction,
this deviation is however less prominent than for real data 
\cite[e.g.][]{maughan2007,arnaud2010}.
\\
In \fig\ref{fig:ml}, we display the relation and the best-fit curve. For the purpose of comparison,
we also show the self-similar line, normalized in mass to the same pivot used for the best-fit, 
i.e. $7.5\times10^{14}~h^{-1}\msun$.
\subsubsection{Correlations with $Y_X$}\label{sec:yx_rel}
\begin{figure*}
\centering
\includegraphics[width=0.33\textwidth]{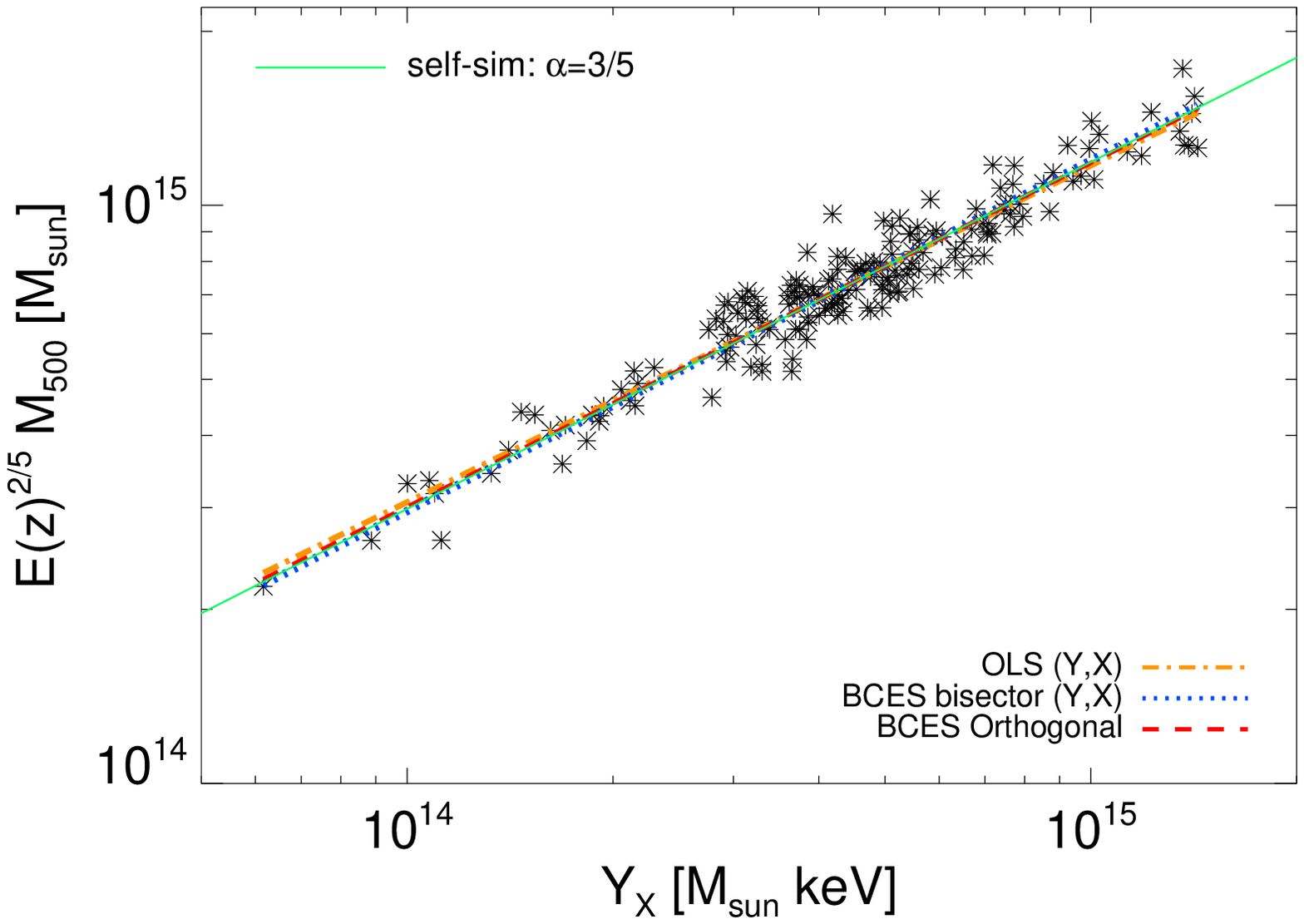}
\includegraphics[width=0.33\textwidth]{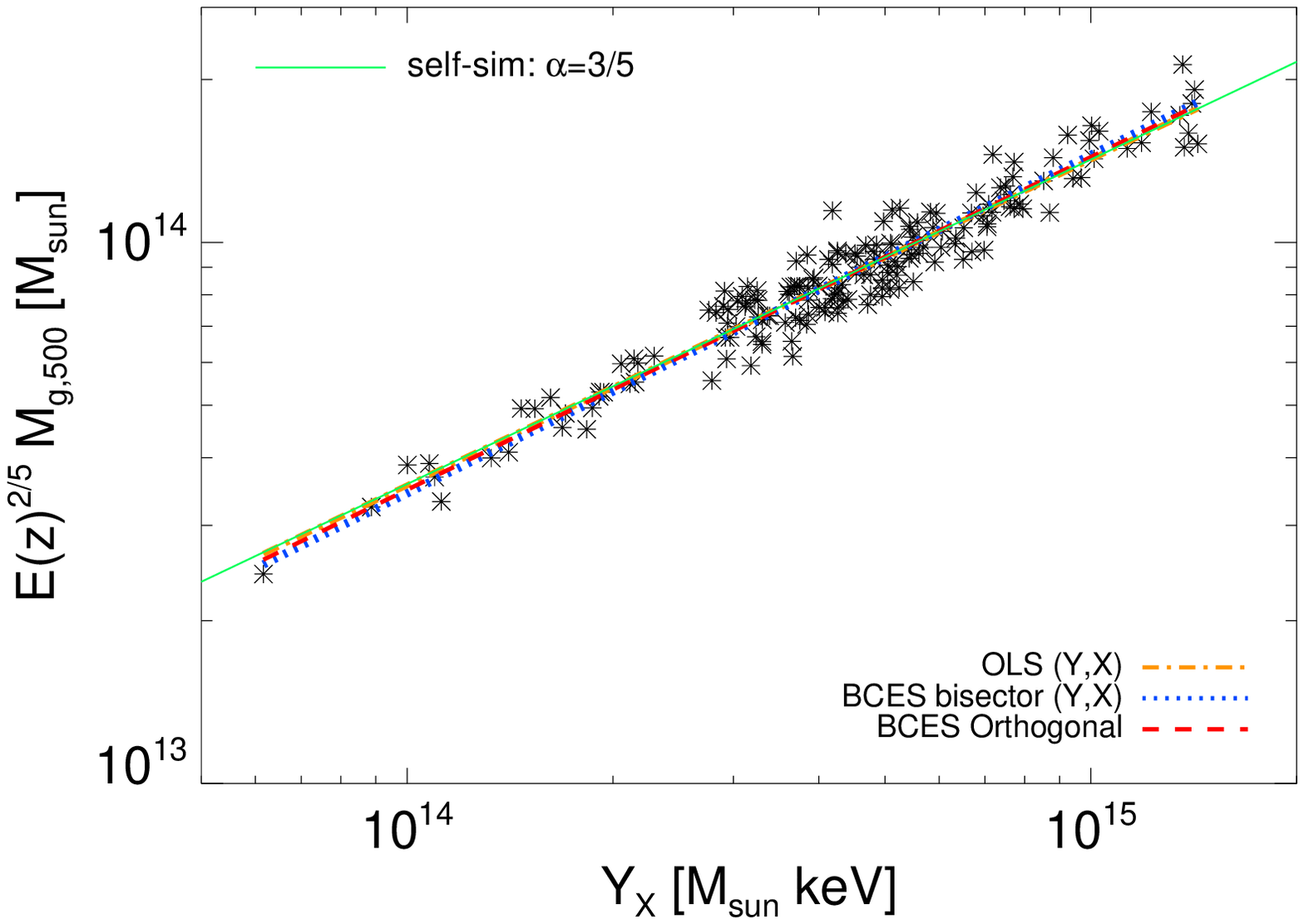}
\includegraphics[width=0.33\textwidth]{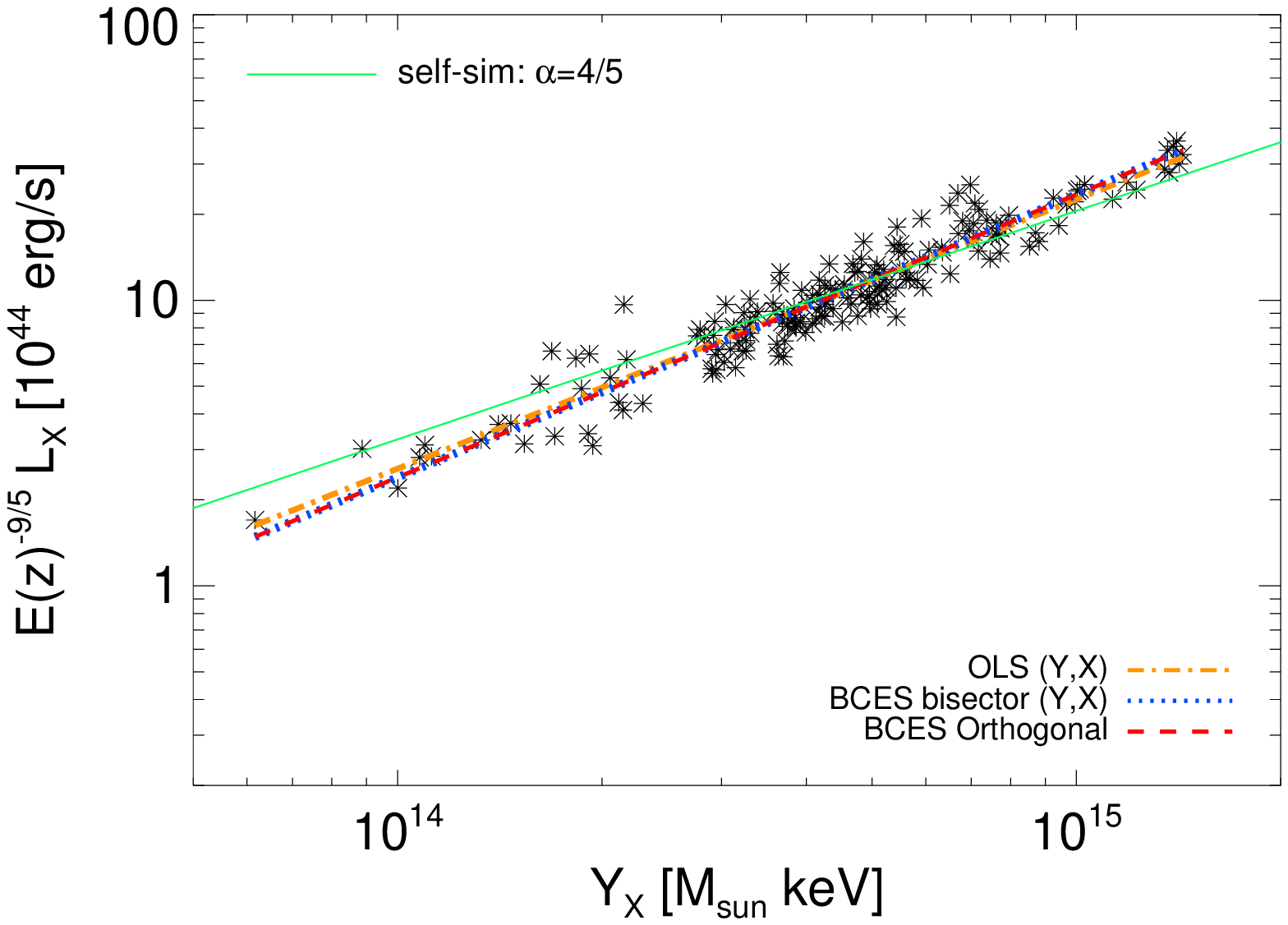}
\caption{$\mfive - Y_X$ (left), $M_{g,500} - Y_X$ (middle) and 
  $L_X - Y_X$ (right) scaling relations. 
  Best-fit lines for the various approaches are reported as in the legend.
  As a comparison, we
  also over-plot the self-similar predicted curve (solid, green), normalized to the
  data at $Y_X=5\times 10^{14} \msun \kev$.} 
\label{fig:yx}
\end{figure*}
Additionally, it is interesting to address the effects of the measured X-ray temperature
with respect to theoretical, intrinsic quantities inferred directly from
the simulations.
In particular, we report on the correlations between 
the total and the gas mass of the clusters, 
enclosed within $\rfive$ 
($\mfive$ and $M_{g,500}$, respectively),
and the $Y_X$ parameter 
\cite[introduced by][]{kravtsov2006}, 
defined as
\begin{equation}\label{eq:yx}
 Y_X = M_{g} T_X.
\end{equation}
$Y_X$ basically quantifies the thermal energy of the ICM,
and we evaluate it for the cluster region within $\rfive$.
\\
The $M_{g,500}-Y_X$ and $\mfive-Y_X$ relations are shown in the 
left and middle panels of \fig\ref{fig:yx}.
Here, we employ the X-ray spectroscopic temperature $T_X$ 
but still use the true total ($\mfive$) and gas ($M_{g,500}$) 
mass of the simulated clusters, with the sole purpose of
calibrating the scaling relations and discerning the effects due to 
the X-ray (mis-)estimate of the ICM temperature.
\\
We confirm that, despite the complexity of the ICM thermal structure, the estimate
of temperature derived from X-ray analysis does not influence majorly the shape of the
relation with mass and rather preserves the tight dependence.
The slope of both relations ($0.60 \pm 0.01$ and $0.58 \pm 0.01$, respectively)
is very close to the self-similar value 
($\alpha_{\rm self-sim}=3/5$) and the scatter in the ${\rm Log}M$ is only about 4 per cent.
\\
The $L_X-Y_X$ scaling relation (right panel in \fig\ref{fig:yx}) 
shows a steeper ($\alpha \sim 0.94 \pm 0.02$) dependence 
than expected from self-similarity ($\alpha_{\rm self-sim}=4/5$).
Even in this case, given the expected good correlation between $Y_X$ and the system mass,
as well as between temperature and mass, we can ascribe both the larger scatter 
($\sigma_{{\rm Log}L_X}\sim 0.08$, i.e. a factor of $\sim 2$ larger than 
in $M_{g,500}-Y_X$ and $\mfive-Y_X$) and the deviation from
the theoretical self-similar model to the $L_X$ observable.
Higher values of luminosity for a certain mass can also
be affected by the choice of considering the whole region
within $\rfive$, not excluding the innermost part.
\\
Indeed, we find an overall good agreement if compared to 
similar observational analyses
\cite[see, e.g.,][]{maughan2007,pratt2009}.
Nonetheless, this MUSIC-2 sub-sample, sampling the most massive clusters, 
shows a behaviour slightly closer to self-similarity 
with respect to the observations.
\subsubsection{A pure X-ray scaling relation: $L_X-T_X$}\label{sec:lt_rel}
\begin{figure*}
\centering
\includegraphics[width=0.33\textwidth]{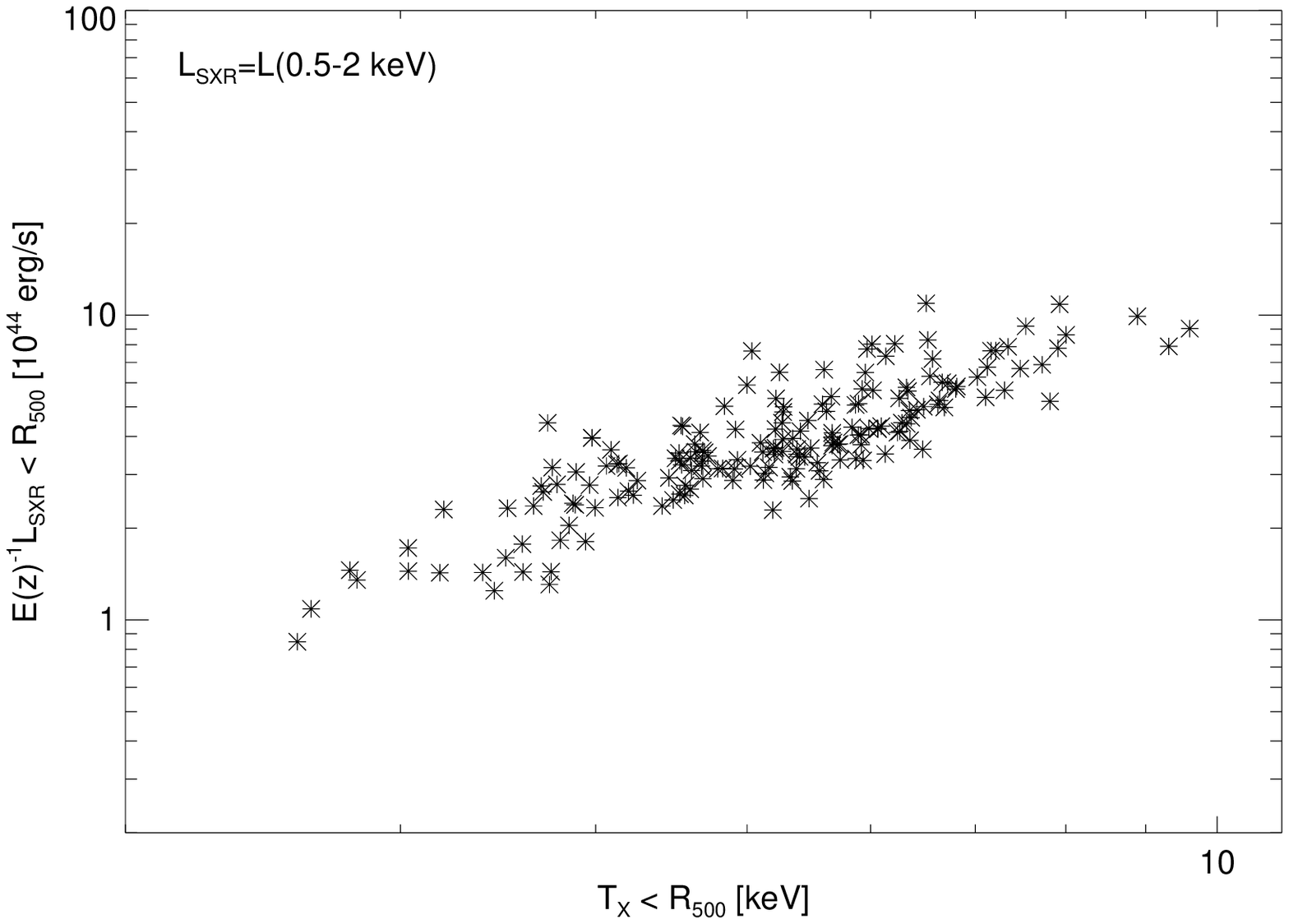}
\includegraphics[width=0.33\textwidth]{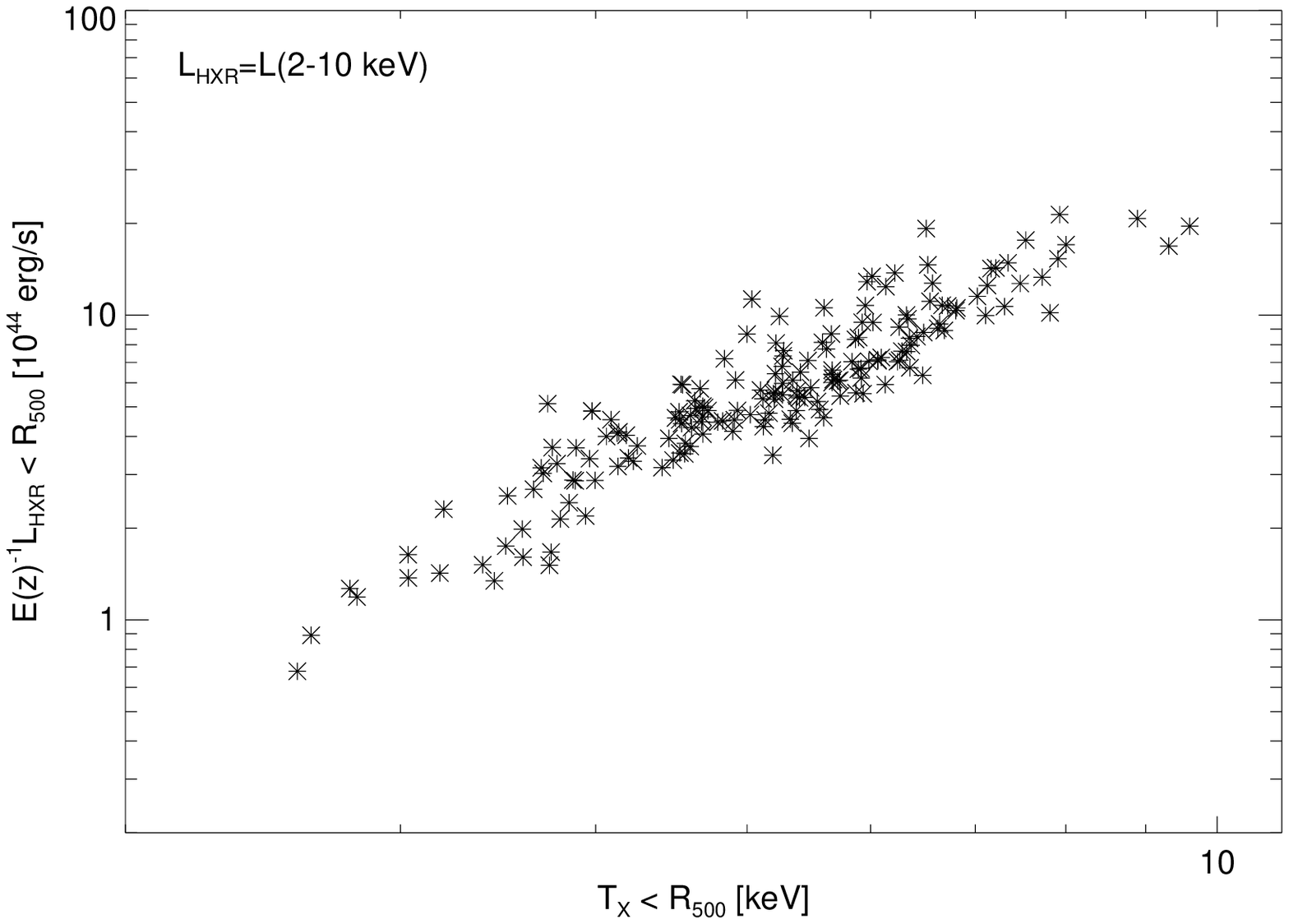}
\includegraphics[width=0.33\textwidth]{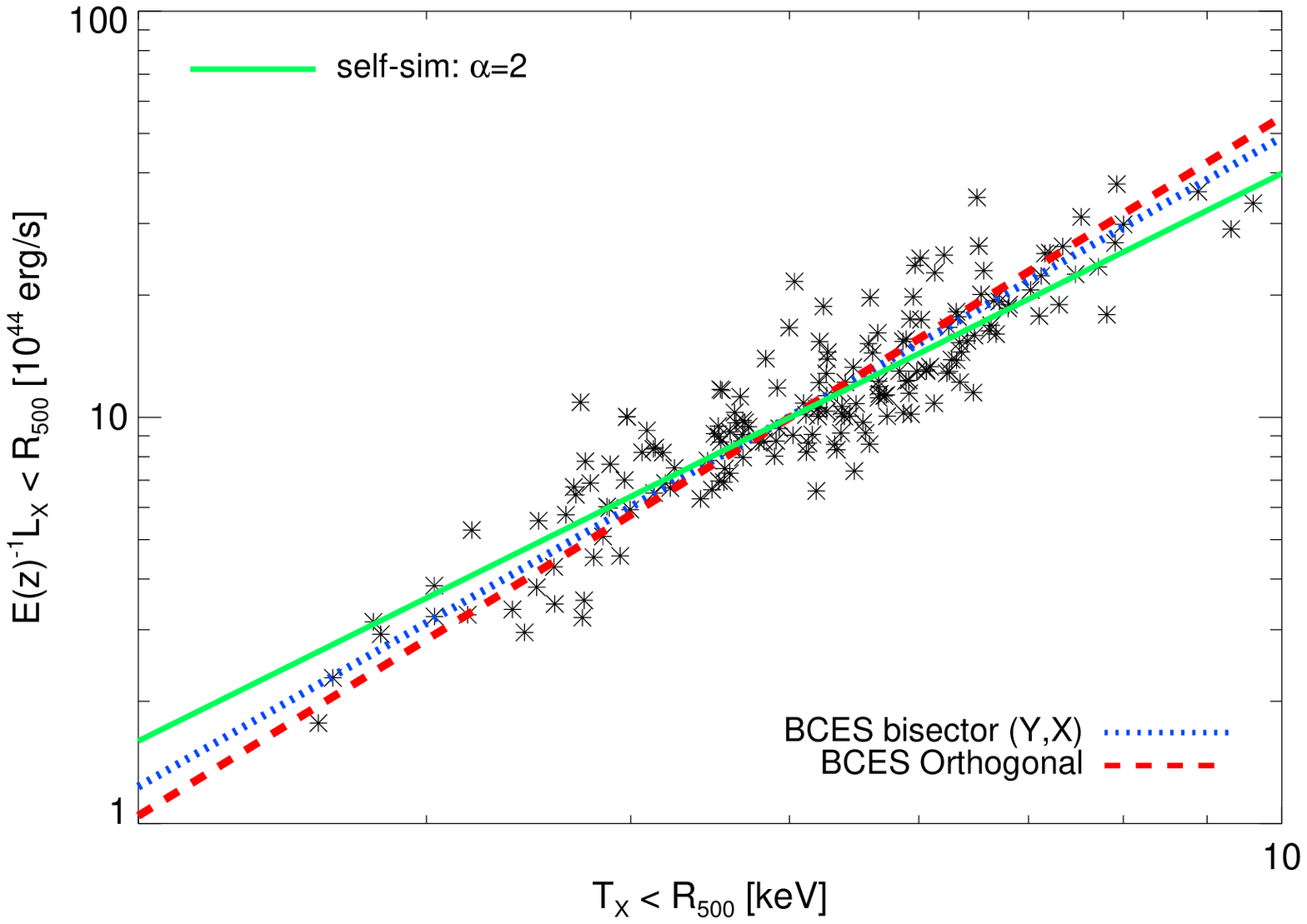}
\caption{Luminosity--temperature relation for the soft (left
  panel), hard (middle panel) and ``bolometric'' (right panel) X-ray band. 
  {\it Right panel:} the best-fit lines for the
  Bisector and Orthogonal methods are marked as in the legend, as well as
  the self-similar curve (green, solid), normalized to the data at the pivot 
  temperature $5 \kev$.} 
\label{fig:ltrel}
\end{figure*}

As a further step, we investigate the relation between X-ray luminosity and spectroscopic
temperature, within $\rfive$ (projected radius), for the sample of clusters.
\\
As described in \sec\ref{sub:lum}, the X-ray luminosity has been
obtained from the best-fit of the synthetic Chandra (ACIS-S) spectra
generated with the PHOX simulator, 
for: 0.5-2~$\kev$ (soft X-ray band, SXR),
2-10~$\kev$ (hard X-ray band, HXR) and for the total band covered by
the ACIS-S detector response (``bolometric'' X-ray luminosity).
The $L_X-T_X$ for the three aforementioned energy bands is shown in 
\fig\ref{fig:ltrel}\footnote{ In the figures, 
the luminosity is always reported in units of
$10^{44}\ergs{}$ and the temperature in $\kev$. 
We note that errors obtained form spectral fitting are here
very small, given the relatively good statistics of photon counts. Hence,
for the clarity of the figure, we decide not to show them.}.\\
In order to study the luminosity--temperature scaling law more quantitatively,
we perform a linear fit to the $L_X-T_X$ relation\footnote{The relation
  considered for the best-fit analysis involves the X-ray
  ``bolometric'' luminosity, as in
  \fig\ref{fig:ltrel}.} in the Log-Log plane, in order to find the slope, $\alpha$, and
normalization, $C$, of the best-fit relation.
\\
Here the functional form in \eq\ref{eq:scal_rel_log} reads:
\begin{equation}\label{eq:lt_fit}
{\rm Log} (E(z)^{-1}\,L_X) = {\rm Log}(C) + \alpha{\rm Log} (T_X/T_0),
\end{equation}
where $L_X$ is given in units of $10^{44}\ergs$, as well as the
normalization $C$, and we assume $T_0=5\kev{}$.
\\
For the purpose of comparison, we recall the self-similar
expectation for the luminosity--temperature relation:
\begin{equation}\label{eq:lt}
E(z)^{-1}L\propto T^2.
\end{equation}
The resulting slope and normalization are sensitive to the method adopted
to minimize the residuals, so that a cautious interpretation of the
two observables involved in the relations is recommended.
We notice that, in the particular case of the $L_X-T_X$ relation,
we might interpret the luminosity as the ``dependent'' variable and
the temperature as the ``independent'' one, being the latter closely
related to the total mass of the cluster (see \fig\ref{fig:tm_rel} and
discussion in \sec\ref{sec:mt}) and therefore tracing an intrinsic
property of the system.
Under this assumption, the standard OLS method, minimizing only the residuals in
the luminosity, would suggest a shallow relation, quite close to the
self-similar prediction, with $\alpha=2.08\pm 0.07$.
\\
Nevertheless, a more careful approach to find the best-fit slope
consists in a minimization procedure that accounts for both the residuals 
in $L_X$ and $T_X$, without any stringent assumption on which variable has
to be treated as (in)dependent.
Therefore, we apply here the linear regression BCES method, focusing
on the Bisector (Y,X) and Orthogonal modifications
\cite[][]{isobe1990,akritas1996}.
Both methods are in fact robust estimators of the best-fit slope and
provide us with more reliable results than the OLS approach. 
\\
For the present analysis, the best-fit values, with
their 1-$\sigma$ errors, and the scatter (see \eq\ref{eq:scatter}) are 
listed in \tab\ref{tab:lt_rel}.
We highlight that the scatter of the relation for this MUSIC-2
subsample is lower ($\sigma_{{\rm Log L_X}} \sim 0.11$), with respect
to what is usually found when the relation is calculated for a density contrast $\Delta=500$ 
\cite[see both numerical and observational studies on the $L_X-T_X$ relation, e.g.][]{ettori2004_obs,maughan2007,pratt2009,biffi2013,biffiAN}.
\\
Between the two methods employed, as already pointed out in previous works,
the Orthogonal BCES provides a steeper best-fit relation to the data
than the Bisector (Y,X) method.
Nonetheless, the slope of the relation for the MUSIC-2 clusters is found to be
in general shallower, and in slightly better agreement to the
self-similar prediction, than often observed at cluster scales
\cite[e.g.][]{white1997,markevitch1998,arnaud1999,ikebe2002,ettori2004_obs,maughan2007,zhang2008,pratt2009}.
Here, we notice that the relation is mainly constrained in the
high-temperature envelope of the $L_X-T_X$ plane (for all the
clusters $T_X>2\kev$) and we would rather expect a larger statistics in
the low-temperature region to introduce a larger deviation from the
theoretical expectation. 
Indeed, a remarkable steepening of the relation is observed especially
at 
galaxy-group scales (or equivalently, for systems with temperatures $<2-3\kev$),
even though this is still a very debated issue \cite[e.g.][]{ettori2004,eckmiller2011}. 
In agreement with our findings, previous studies indicate a possibly shallower slope that
approaches the self-similar expectation for very hot
systems \cite[e.g.][]{eckmiller2011}, 
which would be the case for the majority of our clusters
(see, e.g., \fig\ref{fig:ltrel}, where only 4 objects
have $2\kev<T_X<3\kev$).
\\
In general, the limitations related to the description of the baryonic physics
acting in their central region strongly affect the
final appearance of simulated clusters, 
which still fail to match some observational features.
Among these, the $L_X-T_X$ surely represents a critical issue. 
Certainly, an incomplete description of feedback processes (e.g. from AGN),
turbulence \cite[see, e.g.][]{vazza2009} and galaxy evolution can weaken the departure of
simulated clusters from the theoretical, gravity-dominated scenario
and consequently augment the gap between cluster simulations and
observations \cite[e.g.][for a recent review]{borgani2011}.
\\
In the case of observed galaxy clusters, in fact, the deviation
from the theoretical expectation is often definitely more striking than
what is observed for our simulated sample.
\\
As a final remark, we note that the steepening of the MUSIC $L_X-T_X$
relation can also possibly point to the combination of two effects:
the effect of temperature under-estimation
and the possible over-estimation of luminosity, artificially
increased in the center because of the incomplete feedback treatment.
To this, also the choice of not removing the core from the current analysis 
can additionally contribute and further investigations 
in this direction will be worth a separate, dedicated study.
\subsubsection{Comparison to SZ-derived properties}\label{sec:x-sz}
We study here correlations between properties of clusters derived from 
both X-ray synthetic observations and estimates of the SZ signal,
in order to build mixed scaling relations
for the sample of MUSIC-2 clusters analysed.
Both approaches, in fact, allow us to investigate in a complementary
way the properties of the hot diffuse ICM and to assess the effects of
the baryonic physical processes on the resulting global features
\cite[e.g.][]{mccarthy2003,dasilva2004,bonamente2006,bonamente2008,morandi2007,arnaud2010,melin2011}.
\\
Regarding the SZ effect, of which we only focus here on the thermal component,
we recall that the Comptonization $y$ parameter 
towards a direction in the sky is defined as:
\begin{equation}
y = \int n_e\, \frac{k_BT_e}{m_ec^2}\, \sigma_T {\rm d}l.
\end{equation}
A more interesting quantity, however, is the integrated Comptonization $Y_{SZ}$
parameter, which expresses no more a local property but
rather describes the global status of the cluster, within a region
with a certain density contrast, e.g. $<\rfive$.
As for the estimation of X-ray properties,
such global quantity is therefore
less dependent on the specific modelling of the ICM distribution.
\\
This is given by:
\begin{equation}\label{eq:Ysz}
Y_{SZ} \equiv \int_{\Omega} y~{\rm d}\Omega = D_A^{-2}
\frac{k_B\sigma_T}{m_ec^2}\int_0^\infty {\rm d}l \int _An_eT_e{\rm d}A
\end{equation}
where
$n_e$ and
$T_e$ are electron density and temperature in the ICM, 
$D_A$ is the angular-diameter distance to the cluster and the
integration is performed along the line of sight (${\rm d}l$ is the
distance element along the l.o.s.) and over
the solid angle ($\Omega$) subtending the projected area ($A$) of the
cluster on the sky.
The other constants appearing in \eq\ref{eq:Ysz} are: the Boltzmann
constant, $k_B$, the Thompson cross-section, $\sigma_T$, the speed of
light, $c$, and the rest mass of the electron, $m_e$.
\\
Simulated maps of the Comptonization parameter $y$ have been generated
for the MUSIC clusters, 
and from them we evaluate the integrated $Y_{SZ}$ 
within a radius $\rfive$, i.e. $Y_{SZ,500}$
\cite[see][for further details]{sembolini2012}.
Throughout our analysis, as commonly done, 
we consider the quantity $Y_{SZ,500}\,D_A^2$
and re-name it as: 
\begin{equation}\label{eq:Ysz_da2}
 Y_{SZ,500}\,D_A^2 \longrightarrow Y_{SZ},
\end{equation}
simply referring to $Y_{SZ}$ hereafter.
\\
As $Y_{SZ}$ has proved to be a good, low-scatter mass proxy
\cite[confirmed also by][for the MUSIC-2 data-set]{sembolini2012},
it is interesting to explore its relationship with other global 
cluster properties commonly observed.
Precisely,
our principal aim is to confront this SZ-derived quantity describing 
the ICM to global properties obtained instead from the X-ray analysis.
\subsubsection{The $Y_{SZ}-T_X$ and $Y_{SZ}-L_X$ relations}\label{sec:ylt_rel}
\begin{figure*}
\centering
\includegraphics[width=0.49\textwidth]{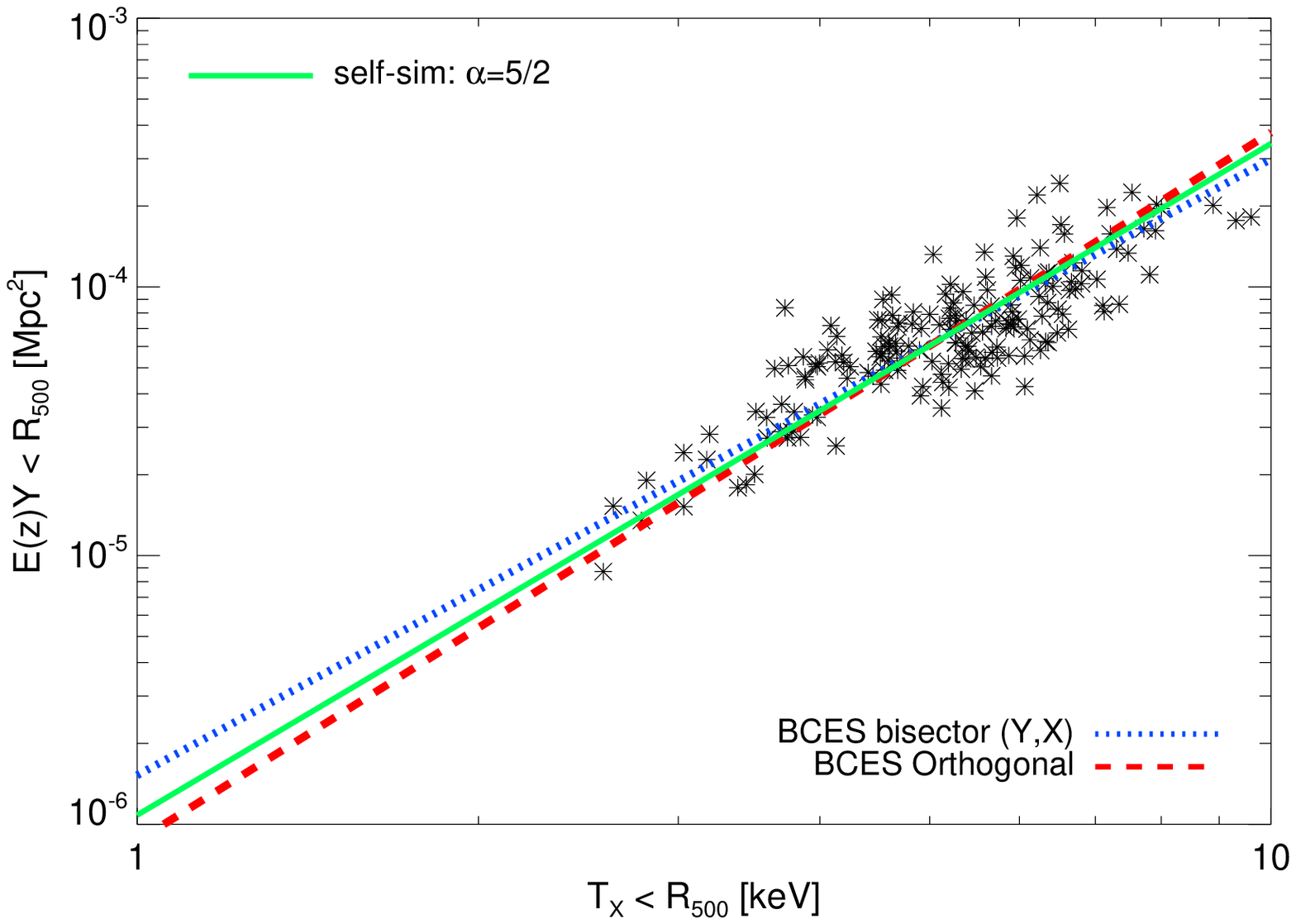}
\includegraphics[width=0.49\textwidth]{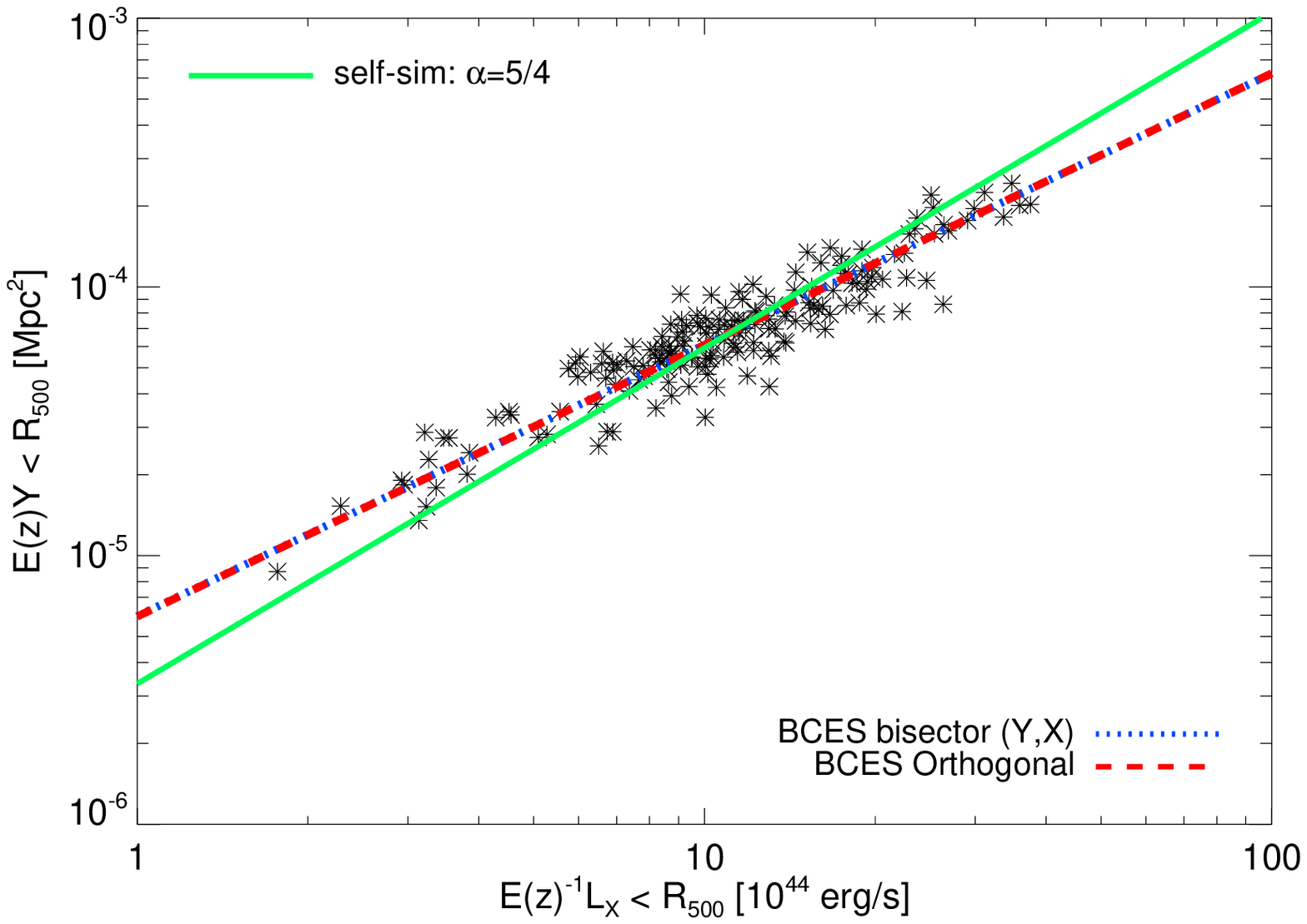}
\caption{Relations between integrated Comptonization parameter and
  X-ray temperature and luminosity: $Y_{SZ}-T_X$ (left panel)
  and $Y_{SZ}-L_X$ (right panel). 
  Best-fit lines for the Bisector and 
  Orthogonal approaches are reported as in the legend.
  Self-similar curves are reported for comparison purposes (green, solid curves),
  normalized to the data at $5 \kev$ and $10^{45}\ergs$ for $T_X$ and $L_X$, respectively.} 
\label{fig:ylt_rel}
\end{figure*}
First, we investigate the relation between the integrated Comptonization
parameter $Y_{SZ}$ (\eq\ref{eq:Ysz}, with definition \ref{eq:Ysz_da2}) and the X-ray temperature and ``bolometric'' luminosity
($T_X$ and $L_X$) within the projected $\rfive$ 
\cite[see, e.g.,][]{dasilva2004,arnaud2007,morandi2007,melin2011}.
\\
The self-similar scaling of these quantities
predicts:
\begin{eqnarray}
  \label{eq:yt}
  E(z)\,Y_{SZ} &\propto& T^{5/2},\\
  \label{eq:yl}
  E(z)\,Y_{SZ} &\propto& (E(z)^{-1}L)^{5/4}.
\end{eqnarray}
In a more realistic picture, the additional complexity of baryonic
processes is most-likely responsible for the deviation from the
theoretical prediction.
Additionally, observational limitations, such as instrumental response, projection effects and
modelling of the data, also play a role in the final shape of reconstructed relations 
and in the discrepancy with theory.
\\
Also in this case, we fit the synthetic data obtained
for the MUSIC-2 clusters assuming a functional form similar to \eq\ref{eq:scal_rel_log}:
\begin{equation}
{\rm Log} (E(z)\,Y_{SZ}) = {\rm Log}(C) + \alpha{\rm Log} ({\rm X}),
\end{equation}
where $Y_{SZ}$ is the SZ integrated Comptonization
parameter given by \eq\ref{eq:Ysz} and re-definition as in \eq\ref{eq:Ysz_da2}, 
and the variable ${\rm X}$ is replaced either with $T_X$
(\eq\ref{eq:yt}) or with $E(z)^{-1}L_X$ (\eq\ref{eq:yl}).
Both the integrated Comptonization parameter $Y_{SZ}$ and the normalization
of the relations $C$ are given in units of $\mpc^2$, while the X-ray
luminosity and temperature are in units of $10^{44}\ergs$ and $\kev$, respectively.
\\
For both these relations, it is not clear which variable 
between the two quantities involved should be treated as (in)dependent
and a simple OLS minimization of the residuals in the Y variable 
would be most likely inappropriate to provide a reliable fit for the slope.
Therefore, in order to correctly approach the problem, we employ also in this case the
Bisector and Orthogonal methods and minimize the residuals of both variables
with respect to the best-fit relation, providing results for both.
\\
We show the scaling relations in \fig\ref{fig:ylt_rel} and report the
best-fit values for slope and normalization, together with their 1-$\sigma$ errors, in \tab\ref{tab:ylt_rel}.
There we also list the scatter (\eq\ref{eq:scatter}) around the best-fit laws 
($\sigma_{{\rm Log}Y}\sim 0.15$ for the $Y_{SZ}-T_X$ relation and $\sim 0.10$ for the $Y_{SZ}-L_X$).
\\
From the values listed in \tab\ref{tab:ylt_rel} for the $Y_{SZ}-T_X$ relation 
we notice that the Orthogonal BCES method converges on 
steeper slopes than the Bisector method,
as in the case of the $L_X-T_X$ scaling law.
In particular, we find that the slope of the $Y_{SZ}-T_X$ relation better agrees with 
the predicted self-similar value than the slope of the $Y_{SZ}-L_X$ one, which is 
equally under-estimated by both Bisector and Orthogonal methods, 
rather suggesting a slope shallower than the self-similar prediction.
Given that $Y_{SZ}$ closely traces the system mass, the deviation of $Y_{SZ}-L_X$
from self-similarity can be mainly related to the X-ray luminosity, which is more sensitive 
to the gas physics and dynamical state than to the temperature, 
as already shown in previous sections. 
\\
Comparing the correlations in \fig\ref{fig:ylt_rel},
we note that these results are consistent with the departure of the $L_X-T_X$ scaling law 
from the self-similar trend previously discussed, which was steeper than theoretically predicted.
\\
Additionally, this behaviour is fairly consistent with other results
in the literature, e.g. with findings by \cite{dasilva2004} obtained
from numerical simulations as well as with observational studies by
\cite{arnaud2007,morandi2007,melin2011,planckXIe}. 
\subsubsection{The $Y_{SZ}-Y_X$ relation}\label{sec:yyx_rel}
\begin{figure}
\includegraphics[width=0.47\textwidth]{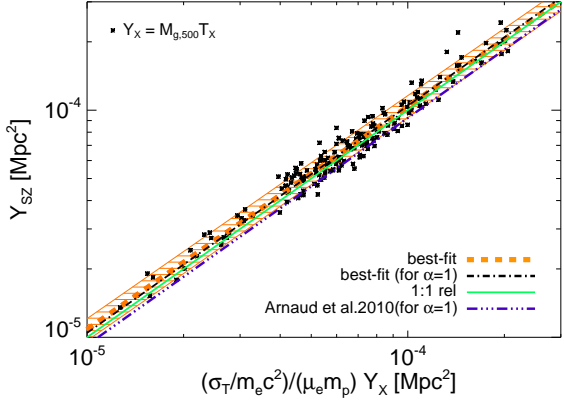}
\caption{Relation between integrated Comptonization parameter $Y_{SZ}$ and the
  X-ray-analog quantity $Y_X=M_{g,500} T_X$. We display in the plot the best-fit relation to the data (orange, dashed line) and the one obtained by fixing the slope to 1 and fitting for the normalization only (black, dot-dashed), the 1:1 (self-similar) relation (green, solid line) and the observational result by Arnaud et al.~(2010) (dot-dot-dot-dashed, purple line). The shaded orange area marks the $5\%$ scatter about the best-fit line.} 
\label{fig:yyx}
\end{figure}

The cluster integrated thermal energy is quantified both by $Y_{SZ}$ and $Y_X$,
with the main difference that the former depends on the mass-weighted temperature of the gas, 
while the latter is rather dependent on the X-ray temperature, resulting more sensitive 
to the lower-entropy gas.
These two integrated quantities are therefore expected to correlate tightly,
and the comparison allows us to test the thermal state of the ICM and the 
differences between the true and the X-ray temperature
\cite[see, e.g.,][]{arnaud2010,andersson2011,fabjan2011,kay2012}.
\\
In order to compare directly $Y_{SZ}$ and $Y_X$, we rescale the latter by the factor
\begin{equation}\label{cszx}
 C_{SZX} = \frac{\sigma_T}{m_e c^2}\frac{1}{\mu_e m_p}=1.43\times 10^{-19} \frac{\mpc^2}{\msun \kev}
\end{equation}
for a mean molecular weight of electrons $\mu_e=1.14$.
\\
In \fig\ref{fig:yyx} we show the $Y_{SZ}-Y_X$ relation, for the sample of 179 MUSIC clusters 
with $Y_X = M^{sim}_{g,500} T_X$ (\eq\ref{eq:yx}). 
By using the true gas mass within $\rfive$, calculated directly from the simulations, 
we explicitly investigate the role of temperature.
\\
Indeed, since in this case no deviation is included because of the X-ray estimation of the total
gas mass, the difference between the 1:1 relation and the best-fit line is substantially
attributed to the mis-estimation of $T_{mw}$ by $T_X$. 
More evidently, the ratio $Y_{SZ}/C_{SZX}Y_X$ can be quantified by the best-fitting normalization 
of the relation when the slope is fixed to one.
\\
An ideal measurement of the true temperature of the clusters would basically
permit to evaluate by $Y_X$ the very same property of the ICM as done via $Y_{SZ}$,
expecting an actual 1:1 correlation.
Dealing with simulated galaxy clusters, this can be tested by employing the true $T_{mw}$
in \eq\ref{eq:yx}. By applying this to our subsample of MUSIC clusters, we confirm this
with very good precision.
\\
When the spectroscopic temperature is instead employed,
for a slope fixed to one in the best-fit, we observe a higher normalization,
$C = 1.05$ (for $Y_{SZ}$ and $Y_X$ both normalized to $5\times 10^{-4}\mpc^2$), 
with a scatter of roughly $5$ per cent.
The deviation from one, $\sim 5\%$, is consistent with the
mean deviation between mass-weighted and X-ray temperatures 
(see \fig\ref{fig:bias} and discussion in \sec\ref{sec:temp}).
Nevertheless, this deviation is also comparable to the scatter in the
relation as, in fact, the $T_{mw}$ and $T_X$ estimates for this MUSIC 
sub-sample are in very good agreement.
\\
Moreover, we note that the employment of the X-ray temperature generates
a larger scatter about the relation, 
although relatively low with respect to
previous (X-ray) relations ($\sigma_{{\rm Log}Y_{SZ}}\sim 0.05$).
This confirms the robustness of $Y_X$ as a mass indicator, despite the 
minor deviations due to $T_X$.
As a comparison, the ideal, reference test for $Y_X = M_{g,500} T_{mw}$ provides a
remarkably tighter correlation, with a scatter smaller than one percent.
\\
We also perform the linear fit to the $Y_{SZ}-C_{SZX}Y_X$ in the Log-Log plane
in order to find the values of the slope and normalization that minimize the residuals,
listed in \tab\ref{tab:yyx} with the scatter in ${\rm Log}Y_{SZ}$.
Even in this case, both slope and normalization are very close to the expected
(self-similar) value of one, with a scatter 
of $\sim 5\%$. 
\\
As marked in \fig\ref{fig:yyx} by the orange, dashed curve and shaded area around it
(which indicates the $5\%$ scatter),
the best-fit relation for MUSIC clusters is still consistent with the expected 
one-to-one relation.
%
	\begin{table*}
	\begin{center}
	\caption[fit]{Best-fit parameters of the $L_X-T_X$ scaling
        relation: slope, normalization and scatter. The relation is normalized at the value of $T_0 = 5\kev$. 
        For comparison, we also report the expected self-similar slope.}
	\label{tab:lt_rel}
	\begin{tabular}{llcccc}
        \hline
	&& ${\bf \alpha_{\rm self-sim}}$ & ${\bf \alpha}$ & ${\bf C}~[10^{44}\ergs]$ & ${\bf \sigma_{{\rm Log}Y_{SZ}}}$ \\
	\hline
	\multirow{2}{*}{$^{\star}{L_X-T_X}$} & BCES Bisector (Y,X) & $2$ & $2.29 \pm  0.07$ & $10.03 \pm 0.19$ & $0.11$\\
                                                          & BCES Orthogonal    & $2$ & $2.46 \pm 0.09$ & $9.98 \pm 0.19$ & $0.11$\\
        \hline
	\end{tabular}
	\end{center}
	\begin{flushleft}
	$^{\star}${\scriptsize The luminosity considered is calculated over the maximum
        energy band defined by the Chandra ACIS-S response, as in \fig\ref{fig:ltrel}.}\\
	\end{flushleft}
%
	\begin{center}
	\caption[fit]{Best-fit parameters of the $Y_{SZ}-T_X$ and $Y_{SZ}-L_X$ scaling
          relations: slope, normalization and scatter. For comparison, we also report the expected self-similar slope.}
	\label{tab:ylt_rel}
	\begin{tabular}{llcccc}
        \hline
	&& ${\bf \alpha_{\rm self-sim}}$ & ${\bf \alpha}$ & ${\bf C}~[10^{-6} \mpc^2]$ & ${\bf \sigma_{{\rm Log}Y_{SZ}}}$ \\
        \hline
	\multirow{2}{*}{${Y_{SZ}-T_X}^{\star}$} & BCES Bisector (Y,X) & $2.5$ & $2.29 \pm  0.09$ & $1.52 \pm 0.22$ & $0.14$\\
                                                             & BCES Orthogonal     & $2.5$ & $2.64 \pm  0.12$ & $0.87 \pm 0.17$ & $0.16$\\
        \hline
	\multirow{2}{*}{${Y_{SZ}-L_X}^{\star}$} & BCES Bisector (Y,X) & $1.25$ & $1.01 \pm 0.03$ & $5.95 \pm 0.40$ & $0.10$\\
                                                             & BCES Orthogonal    & $1.25$ & $1.01 \pm 0.03$ & $5.93 \pm 0.43$ & $0.10$\\
        \hline
	\end{tabular}
	\end{center}
	\begin{flushleft}
	$^{\star}${\scriptsize The luminosity and temperature
          considered are those employed in the $L_X-T_X$ relation (see
          \sec\ref{sec:lt_rel}).}
	\end{flushleft}
	\begin{center}
	\caption[fit]{Best-fit parameters of the $Y_{SZ}-C_{SZX}Y_X$ scaling
        relation: slope, normalization and scatter, with different minimization methods.
        For comparison, we also report the expected self-similar slope.}
	\label{tab:yyx}
	\begin{tabular}{llcccc}
	\hline
	&& ${\bf \alpha_{\rm self-sim}}$ & ${\bf \alpha}$ & ${\bf C^{\dag}}$ & ${\bf \sigma_{{\rm Log}Y_{SZ}}}$ \\
	\hline
	\multirow{3}{*}{$Y_{SZ}-C_{SZX}Y_X$} & OLS (Y,X) & 1 & $0.98 \pm  0.01$ & $1.01 \pm 0.03$ & 0.05 \\
                                             & {BCES Bisector (Y,X)} & 1 & $1.00 \pm 0.01$ & $1.05 \pm 0.03$ & 0.05 \\
                                             & {BCES Orthogonal} & 1 & $1.00 \pm 0.01$ & $1.05 \pm 0.03$ & 0.05 \\ 
	\hline
	\end{tabular}
	\end{center}
	\begin{flushleft}
	$^{\dag}$Both $Y_{SZ}$ and $C_{SZX}Y_X$ are normalized at the pivot point $5\times 10^{-4}\mpc^2$, so that 
        for the slope fixed to one, the normalization expresses directly the ratio $Y_{SZ}/C_{SZX}Y_X$.
	\end{flushleft}
	\end{table*}
\subsubsection{Comparison to observational results}\label{sec:comparison}
Despite the observational-like approach applied to derive X-ray
properties, the MUSIC scaling relations still present some
differences with respect to observational findings.
The aim of this section is to discuss our results, and the level of
agreement with previous observational and numerical studies, given the
strong sensitivity of X-ray cluster properties to the modeling of the
baryonic physics.
To this end, we focus on the mass-temperature and
luminosity-temperature relations, in order to close the circle between
X-ray observables and intrinsic total mass.
\\

\begin{figure}
\centering
\includegraphics[width=0.46\textwidth]{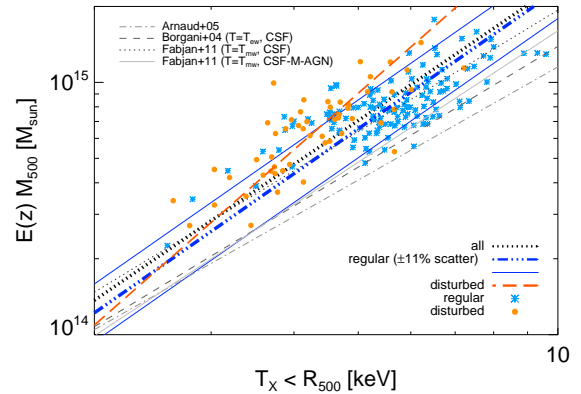}
\caption{Total mass within $\rfive$ as a function of
    temperature. Results are shown considering the
    X-ray spectroscopic temperature ($T_X$), for the regular (cyan asterisks) and
    disturbed (orange circles) sub-samples.  
    Data and best-fit relations from observations by
    \citet{arnaud2005} and from numerical studies by
    \citet{borgani2004} and \citet{fabjan2011} are also shown for comparison.
}
\label{fig:mt2}
\end{figure}

\noindent
$\mathbf{\bullet\,\,\,\mfive-T_X.}$
In the calibration of the mass-temperature relation, the 
estimate of $T_X$ plays a role on the normalization as well. 
In order to investigate this aspect, we show in \fig\ref{fig:mt2} the
inverse relation $\mfive-T_X$, as more commonly presented by several
authors. 
The best-fit curve to the MUSIC data is again fitted, minimizing
the residuals in ${\rm Log}T_X$ (OLS(X$|$Y)), as we consider here the
true mass of the systems.
Consistently with the findings for the $T_X - \mfive$ scaling relation
(\sec\ref{sec:mt}), the slope here is steeper than self-similar.
Moreover, compared to observational data, in particular to the relation by
\cite{arnaud2005}, we find a higher normalization for the MUSIC
sample.
Part of this difference can be explained by the observational
procedure to derive the total mass from the X-rays, which is known
to under-estimate the true dynamical 
mass
of the system
(see early studies by \citealt{evrard1990,evrard1996} and more recent
works by \citealt{rasia2006,nagai2007,piffaretti2008,jeltema2008,lau2009,morandi2010,rasia2012,lau2013}),
i.e. the intrinsic value which is instead used for the MUSIC clusters.
Nevertheless, additional effects must play a role in increasing the
discrepancy, as this still persists when compared to other numerical
works.
Namely, the treatment of the baryonic physics in the MUSIC simulations can
further contribute to this observed off-set, so that,
for a fixed mass, the MUSIC clusters appear to be colder. 
This can be explored, as in \fig\ref{fig:mt2}, 
by comparing the MUSIC relation to simulation
studies by \cite{borgani2004} and \cite{fabjan2011}, 
that also involve the true mass of the systems.
In particular, we focus on the sub-sample of MUSIC regular clusters,
for which we find on average a very small bias between $T_X$ and
$T_{mw}$.
As a consistency check, we compare first against the
results
by \cite{borgani2004} and \cite{fabjan2011} (``CSF''), 
as they
consider the same physical description of the gas as
the MUSIC re-simulations,
basically including cooling and star formation according to the
standard model by
\cite{springel2003}. 
The difference with respect to the former is simply due to the
difference in the temperature definition, which is the
emission-weighted value in their case; 
with respect to
\cite{fabjan2011} (``CSF''), where $T_{mw}$ is used instead, we find
indeed agreement between the two relations, within the scatter.
When the MUSIC data are instead compared to the results by
\cite{fabjan2011} for runs including metal cooling and AGN feedback
(``CSF-M-AGN''), we find a larger, although not
prominent, deviation.
As the mass considered is always the total intrinsic value from the
simulation and $T_X$ in MUSIC clusters is close to the $T_{mw}$
estimate, 
we expect the discrepancy to be mainly due to the different models
included to describe the baryonic processes.
\\

\begin{figure}
\centering
\includegraphics[width=0.46\textwidth]{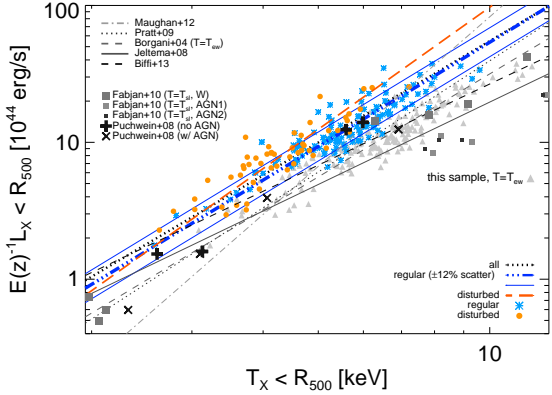}
\caption{$L_X-T_X$ relation. The results for our sample are
    presented for both the regular and disturbed sub-sample of
    clusters (cyan asterisks and orange filled circles, respectively).
    Values corresponding to the
    emission-weighted temperature ($T_{ew}$, light-grey filled
    triangles) are also shown for comparison, as well as observational
    data by \citet{maughan2012} and \citet{pratt2009} and 
    numerical findings by \citet{borgani2004}, \citet{puchwein2008},
    \citet{jeltema2008}, \citet{fabjan2010} and \citet{biffi2013}.
}
\label{fig:ltrel2}
\end{figure}

\noindent
$\mathbf{\bullet\,\,\,\L_X-T_X.}$
Dealing with X-ray properties, the other fundamental quantity taken
into account is the luminosity and its relationship with temperature.
While the steepening of the MUSIC $L_X-T_X$ relation seems
consistent, albeit weaker, with observational results, the
normalization is higher than observed.
In this case, even though some differences between the approach adopted with PHOX and
other observational procedures exist,
the limitations due to the treatment of the baryonic
physics are likely to play a more significant role.
\\
In fact, the lack of an efficient way to remove the hot-phase gas
basically increases the amount of X-ray emitting ICM.
This would be mitigated by the inclusion of AGN feedback, 
although the stronger effects are expected to be particularly important
at group scales, while massive clusters like those in our sample
($T_X>2\kev$) are generally less dramatically affected (as shown by
both \cite{puchwein2008} and \cite{fabjan2010}, despite the different
implementations used). 
In \fig\ref{fig:ltrel2} we show the luminosity-temperature relation
for the two sub-samples of
regular and disturbed clusters. For comparison we also report
observational data by \cite{pratt2009,maughan2012} and results from
numerical studies by
\cite{borgani2004,jeltema2008,puchwein2008,fabjan2010,biffi2013}.
In order to minimize the effects due to X-ray temperature bias, we
specifically focus on regular MUSIC clusters 
(for which $\alpha =
2.53\pm 0.11$ and $C = 9.32 \pm 0.21 10^{44}\ergs$).
\\
Comparing, we note from \fig\ref{fig:ltrel2} 
that for a given temperature the MUSIC clusters appear to be
generally more luminous.
On average, observations \cite[as][]{pratt2009,maughan2012} predict a
luminosity of roughly $6-7 \times 10^{44}\ergs$, at $T_0=5\kev$, while
we find a normalization higher by a factor of $\sim 20-30\%$.
In fact, the inefficient feedback in the center can cause
higher $L_X$.
With respect to those observational works, the difference can
be due in part also to the choice of not removing the core from the
current analysis, where the over-cooling can affect the cluster
central region. 
\\ 
With respect to simulation works employing a spectroscopic
temperature,
as in the numerical studies by \cite{jeltema2008},
\cite{puchwein2008} (run with AGN feedback) and
\cite{biffi2013}, or a spectroscopic-like estimate, as in \cite{fabjan2010} (where the
authors use instead $T_{sl}$), 
the different normalization of the MUSIC relation 
must be mainly related to the missing treatment of AGN feedback
or proper metal cooling. 
In fact, our findings are obviously consistent 
with simulations
accounting for similar models of the baryonic processes
\cite[as for the run withough AGN feedback by][]{puchwein2008}.
This actually explains the diverse level of agreement between the MUSIC
clusters and the results by \cite{puchwein2008} (``w/ AGN'' run) and \cite{fabjan2010}
(``AGN1'' and ``AGN2'' runs),
despite they both account for AGN feedback mechanisms.
\\
Instead, the divergence from the best-fit relation by
\cite{borgani2004} might have a different origin.
Notwithstanding the very similar modeling of the gas physics,
the authors adopt there a different definition of the ICM temperature,
namely the emission-weighted estimate, which brings
the simulated relation closer to both observed results and more complete sets of
hydrodynamical simulations \cite[see][]{jeltema2008,fabjan2010,biffi2013},
predicting a luminosity of $\sim 5.6 \times 10^{44}\ergs$ for $T_0=5\kev$. 
As also confirmed by our analysis, in fact, $T_{ew}$ has been found \cite[e.g.][]{mazzotta2004} to
over-estimate the spectroscopic temperature, which is the one adopted here
instead. Similarly, the use of $T_{ew}$ instead of $T_X$ for the
MUSIC clusters would also provide a lower normalization 
($C = 5.87 \pm 0.15 10^{44}\ergs$) and a better
agreement 
(see light-grey symbols in \fig\ref{fig:ltrel2}).
\\
Additionally, we remark that also numerical resolution can affect
the resulting $L_X$, which can be under-estimated in less
resolved clusters.
Instead, the results tend to reach stability for
increasing resolution \cite[see, for instance,][]{valdarnini2002}. 
Hence, given the similar physical models treated, 
the lower normalization of the clusters in \cite{borgani2004}
can also be partially caused by their lower resolution
with respect to the MUSIC re-simulations.
\\

\noindent The two relations shown in 
\fig\ref{fig:mt2} and \fig\ref{fig:ltrel2} also provide the case to discuss the
behaviour of the regular and disturbed cluster sub-samples.
The two groups of objects clearly occupy different regions of the
relations, having regular clusters on average higher temperatures.
Calculating the two best-fit curves for the two sub-samples
separately, we generally find that 
disturbed clusters provide steeper relations
($\alpha_{M-T}=\pm $ and $\alpha_{L-T}=2.96\pm 0.21$) with respect to
regular objects ($\alpha_{M-T}=\pm $ and $\alpha_{L-T}=2.53\pm 0.11$).
Moreover, 
the disagreement with previous observational and simulations studies
is less significant for the regular sub-set of MUSIC clusters.
\\
The very high statistics of our analysis also provides the case for
studying and constraining the scatter of the relations with very good
precision, despite the level of agreement in terms of slope and
normalization.
In fact, the scatter of the $L_X-T_X$, in particular, is usually significantly
larger in real data 
\cite[up to $50-70$ per cent, as
in][]{pratt2009,maughan2012} than for the MUSIC clusters, where 
instead $\sigma_{{\rm Log}L_{X}}\sim 0.11$.
Considering the two sub-sets separately, the scatter about the
best-fit
relation is slightly different, indicating
a tighter correlation in the first case and a more dispersed relation in
the other, with 
$\sigma_{{\rm Log}L_{X}}\sim 0.10$ (marked by the two solid blue lines in \fig\ref{fig:ltrel2})
and
$\sigma_{{\rm Log}L_{X}}\sim 0.12$,
respectively.
Similarly, while the scatter of the $\mfive-T_X$ relation
is $\sigma_{{\rm Log} M}\sim 0.10$ for the whole sample, the dispersion in ${\rm Log} M$
is found to be smaller for regular objects
($\sigma_{{\rm Log} M}\sim 0.09$, marked by the two solid blue lines in \fig\ref{fig:mt2}) 
and larger for the disturbed ones ($\sigma_{{\rm Log} M}\sim 0.11$). 
\section{Conclusions}\label{sec:discussion}
This analysis presents results on the largest sample of
high-resolution, simulated galaxy clusters ever
analysed with observational approach by means of X-ray synthetic
observations.
\\
Thanks to the large MUSIC-2 data-set, we could obtain a complete volume-limited
sample of re-simulated cluster-like objects.
Out of these, we select a sub-sample of 179 massive haloes at $z=0.11$,
comprising those matching the mass completeness ($M_{vir} > 7.5\times10^{14}h^{-1}\msun$)
at the considered redshift ($z=0.11$), but also extending to all 
the progenitors of the systems with $M_{vir} > 8.5\times10^{14}h^{-1}\msun$ at $z=0$.
%
%
Although restricted to a smaller sub-set, 
our principal goal with this work is to extend the analysis 
on the MUSIC-2 clusters \cite[][]{sembolini2012}
by addressing their X-ray observable properties and scaling relations.
\\
For all the selected objects, we generated ideal X-ray
photon emission \cite[by means of the code PHOX,][]{biffi2012_1} on the
base of the gas thermal properties provided by the original
hydrodynamical simulation. From regions up to $\rfive$ 
centered on each cluster we then obtained Chandra synthetic
observations that provided us with global X-ray properties, 
such as temperature and luminosity (\sec\ref{sec:x-obs}).
\\
First, we investigated the
bias between the spectroscopic
temperature measured from the synthetic spectra ($T_X$) and the theoretical
estimates calculated directly from the simulation.
\\
For the MUSIC sub-sample analysed,
$T_X$ is on average lower then the true, mass-weighted value $T_{mw}$. 
While this is fairly consistent with studies by \cite{mathiesen2001} 
and \cite{kay2008,kay2012},
we observe some tension with X-ray mock studies of simulated clusters
by, e.g., \cite{nagai2007} and \cite{piffaretti2008}.
This discrepancy can be ascribed to the multi-phase thermal structure of the ICM, 
whose temperature distribution plays an important role in the determination
of the global temperature
\cite[see also the detailed discussion in][]{mazzotta2004}.
Indeed, a two-component model would improve the description of 
the multi-temperature structure and provide a better spectral fit, 
albeit with a resulting, evident over-estimation of the true 
temperature by the hotter component of the two.
For the generally massive systems considered, this would generate
a consequent non-negligible bias.
Therefore, we still considered results from the single-temperature,
keeping in mind the tendency by $T_X$ to a mild, average under-estimation 
of
$T_{mw}$.
This difference is very
low in our estimates (roughly $5$ per cent, despite some dispersion; 
see \fig\ref{fig:bias}) 
and we confirm an overall good correlation between
$T_X$, within the projected $\rfive$, and $\mfive$
(\fig\ref{fig:tm_rel}).
\\
The bias between $T_X$ and $T_{mw}$ is also showing
some dependence on the level of dynamical
disturbance of the cluster, quantified by the displacement between the
system center of mass and peak of density. Specifically, regular
clusters show an average 
bias which is consistent with zero (basically dominating the result for the
entire sample), while 
the disturbed sub-set presents a more prominent under-estimation of $T_{mw}$ by $T_X$.
\\
The observational-like derivation of the ICM temperature is also useful to investigate
possible bias in the correlation with intrinsic properties
obtained 
directly from the simulation, as in the $T_X-\mfive$ (\sec\ref{sec:mt})
and $Y_{SZ}-T_X$ (\sec\ref{sec:x-sz}) relations.
The effect has also been studied via the $Y_X$, where the observational-like
$T_X$ is combined with the true $M_{g,500}$ (\sec\ref{sec:yx_rel}).\\
We find that:
\begin{itemize}
 \item $T_X-\mfive$ shows a larger scatter when $T_X$ is employed 
rather than $T_{mw}$
and a shallower slope than expected from the
self-similar scaling (see \fig\ref{fig:tm_rel});
 \item $Y_{SZ}-T_X$ is consistent with findings in observational studies, 
and deviations from self-similarity are less significant than for $Y_{SZ}-L_X$
(\sec\ref{sec:x-sz});
 \item correlations between $Y_X$ and gas or total mass indicate a slope very
close to the self-similar value (\fig\ref{fig:yx});
 \item the employment of X-ray temperature only affects scatter and normalization of 
$M_{g,500}-Y_X$ and $\mfive-Y_X$.
\end{itemize}
Similar considerations can be drawn when the $Y_{SZ}-Y_X$ scaling law is explored
(\sec\ref{sec:yyx_rel}). 
Here the normalization is slightly higher than one, albeit compatible, 
and the slope is remarkably close to self-similarity 
(see \tab\ref{tab:yyx}).
In this case, also the scatter matches the expectation to be
very low ($\sigma_{{\rm Log}Y_{SZ}}\sim 0.05$),
although larger than in the ideal case of $Y_X=M_{g,500} T_{mw}$
(where its is $\lesssim 1\%$).
\\
Unlike $T_X$, the X-ray luminosity is intrinsically less accurate to
trace mass
as it
is particularly susceptible to the non-gravitational processes governing the gas physics.
In fact, $L_X$ is difficult to model in numerical simulations and, from
observations, it is found to add an intrinsic scatter to scaling relations.
Here, we confirm
that $L_X$ tends to augment the deviation from self-similarity as well as the
scatter in the scaling with other intrinsic properties
(such as total mass, SZ integrated Compton parameter or $Y_X$)
and with
X-ray temperature. 
\\
The $L_X-T_X$ scaling relation for this MUSIC sub-sample
extends the study to a larger set of simulated clusters
with respect to what previously done with simulations, especially
involving a proper generation and derivation of observable X-ray quantities 
\cite[][]{puchwein2008,fabjan2010,biffi2013,biffiAN}.
This relation is relatively easy to construct for real clusters
as well and generally represents a crucial break of self-similarity.
In fact, the observed slope significantly deviates from the self-similar
prediction -- typically $\alpha \sim 2.5-3$ instead of $\alpha_{\rm self-sim}=2$ 
\cite[e.g.][]{white1997,markevitch1998,arnaud1999,ikebe2002,ettori2004_obs,maughan2007,morandi2007,zhang2008,pratt2009,maughan2012}.
MUSIC clusters also show a steeper slope than expected, albeit shallower
than in real observations (see \tab\ref{tab:lt_rel}),
when the residuals are minimized for both variables. 
Despite the possible deviations in slope and
normalization, 
the increased statistics of this sample
allows us to precisely estimate
the scatter of the relation, 
which is found to be only $\sim 10\%$ in ${\rm Log}L_{X}$.
\\
From the relations explored, we conclude that 
the interpretation of observational data and comparison to theoretical predictions 
can certainly benefit from the 
observational-like approach.
In fact, a more faithful comparison is possible
even when no
additional complications related to the analysis of real data
(e.g. background subtraction or spacial changes of the effective area) 
are included.
This is especially true for the slope of the relations, which deviates
from self-similarity in a similar way as in observational data.
\\
Differently, the amplitude of the scaling relations is more
sensitive to the accuracy of the physical description 
adopted in hydrodynamical simulations to model the
baryonic processes. 
\\
In fact, the normalization of MUSIC scaling laws shows more tension
with observational findings.
We discuss this and the comparison to other simulation works for the
$\mfive-T_X$ and $L_X-T_X$ relations, which represent the two main
steps to go from X-ray ICM properties to total mass,
via scaling relations. 
Especially in the $L_X-T_X$ case we find that the normalization for
MUSIC clusters is higher than both observations and more complete simulations.
Inefficient cooling and feedback mechanisms, in fact, interplay and compete to moderately
increasing the X-ray emitting gas in the central part of MUSIC
clusters, thereby augmenting the luminosity and, simultaneously,
reducing the temperature. 
\\
Certainly, more robust mass indicators that are not strongly affected by
non-gravitational processes, such as $Y_X$, can be safely employed
\cite[][]{kravtsov2006,nagai2007,fabjan2011}.
In fact, the low scatter around the MUSIC $Y_{SZ}-Y_X$,
$M_{g,500}-Y_X$ and $\mfive-Y_X$ relations is preserved, even when we
use our observational estimates of the X-ray temperature. 
Moreover, in the specific case of $Y_{SZ}-Y_X$, 
the MUSIC clusters are also fairly compatible with observations.
\\
Nevertheless, we remark that,
given the increasingly detailed observations available with current and up-coming 
X-ray instruments 
(e.g. ASTRO-H and ATHENA+), a more complete modeling of the baryonic
physical processes in simulations is required.
Unavoidably, this also needs to
be combined with a proper observational-like approach to derive X-ray properties.
In this way, it will be possible to 
eventually minimize
the distance between numerical
hydro-simulations and observations, and correctly interpret the
complex, underlying ICM physics.
\section*{Acknowledgments}
The authors would like to thank the anonymous referee for valuable comments 
that helped improving the presentation of our work.
VB acknowledges useful discussions with Robert Suhada and Klaus Dolag.
The MUSIC simulations have been performed in the MareNostrum supercomputer at the 
Barcelona Supercomputer Center, thanks to access time granted by the Red Espa\tilde{n}ola de
Supercomputaci\'on, while the initial conditions have been done at the Munich Leibniz 
Rechenzentrum (LRZ).
FS is supported by the MINECO (Spain) under a Beca de Formaci\'on de Profesorado Universitario 
and by research projects AYA2012-31101, and Consolider Ingenio MULTIDARK CSD2009-00064.
GY acknowledges support from MINECO under research grants AYA2012-31101, FPA2012-34694, 
Consolider Ingenio SyeC CSD2007-0050 and from Comunidad de Madrid under ASTROMADRID 
project (S2009/ESP-1496).
MDP has been supported by funding from the University of Rome Sapienza, 2012 (C26A12T3AJ).

\bibliographystyle{mn2e}
\bibliography{bibl.bib}


\appendix
\section[]{Effects on the temperature estimate of the
  single-temperature spectral fit.}\label{appA} 
\begin{figure}
\includegraphics[width=0.44\textwidth,height=0.21\textheight]{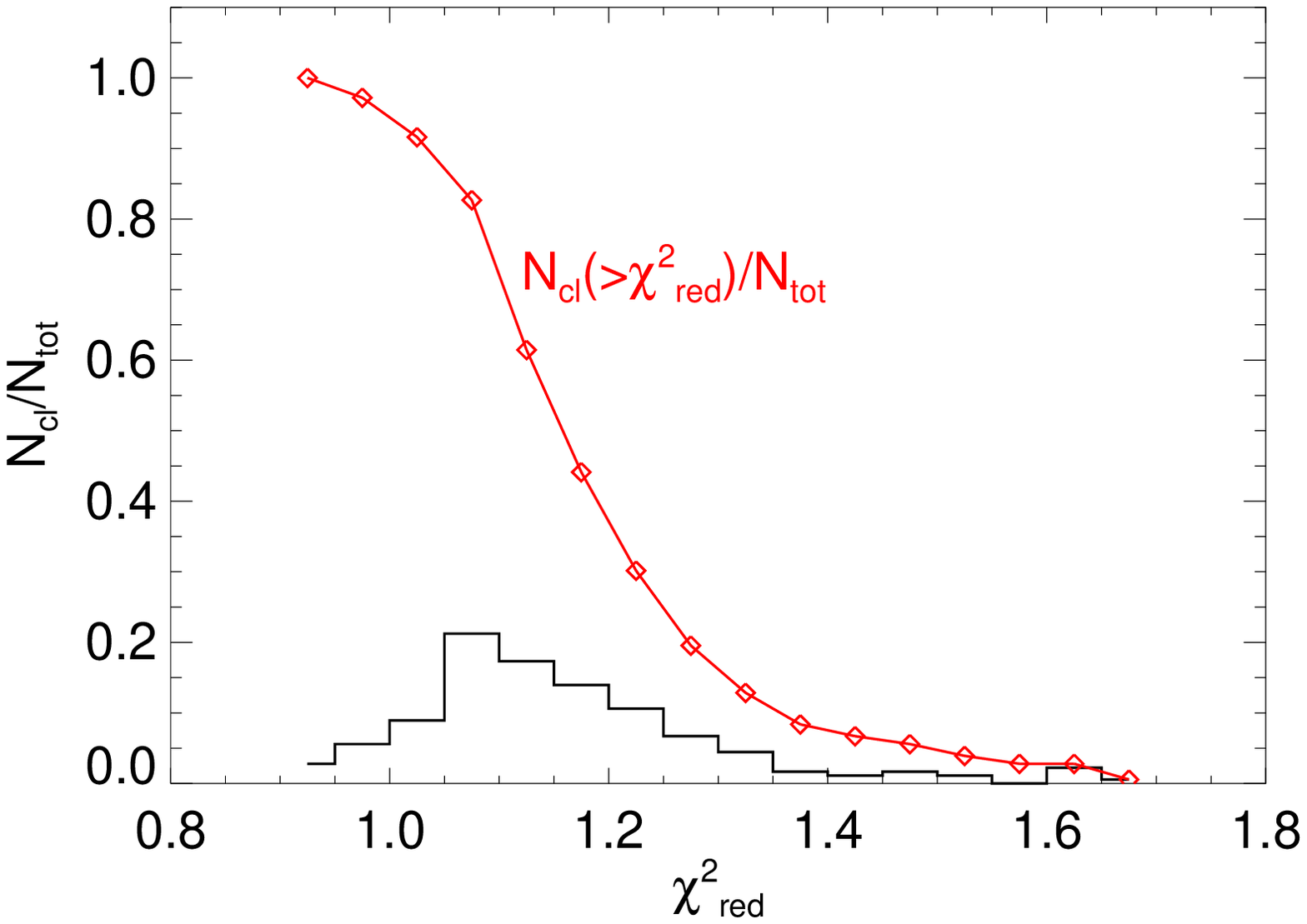}
\caption{Distribution of the reduced-$\chi^2$ values of the
  best-fitting models for the clusters in the sample. Over-plotted is the cumulative
distribution, $N_{cl}(>\chi^2_{red})/N_{tot}$ (solid, red line with diamonds).} 
\label{fig:chi2_fits}
\includegraphics[width=0.44\textwidth,height=0.21\textheight]{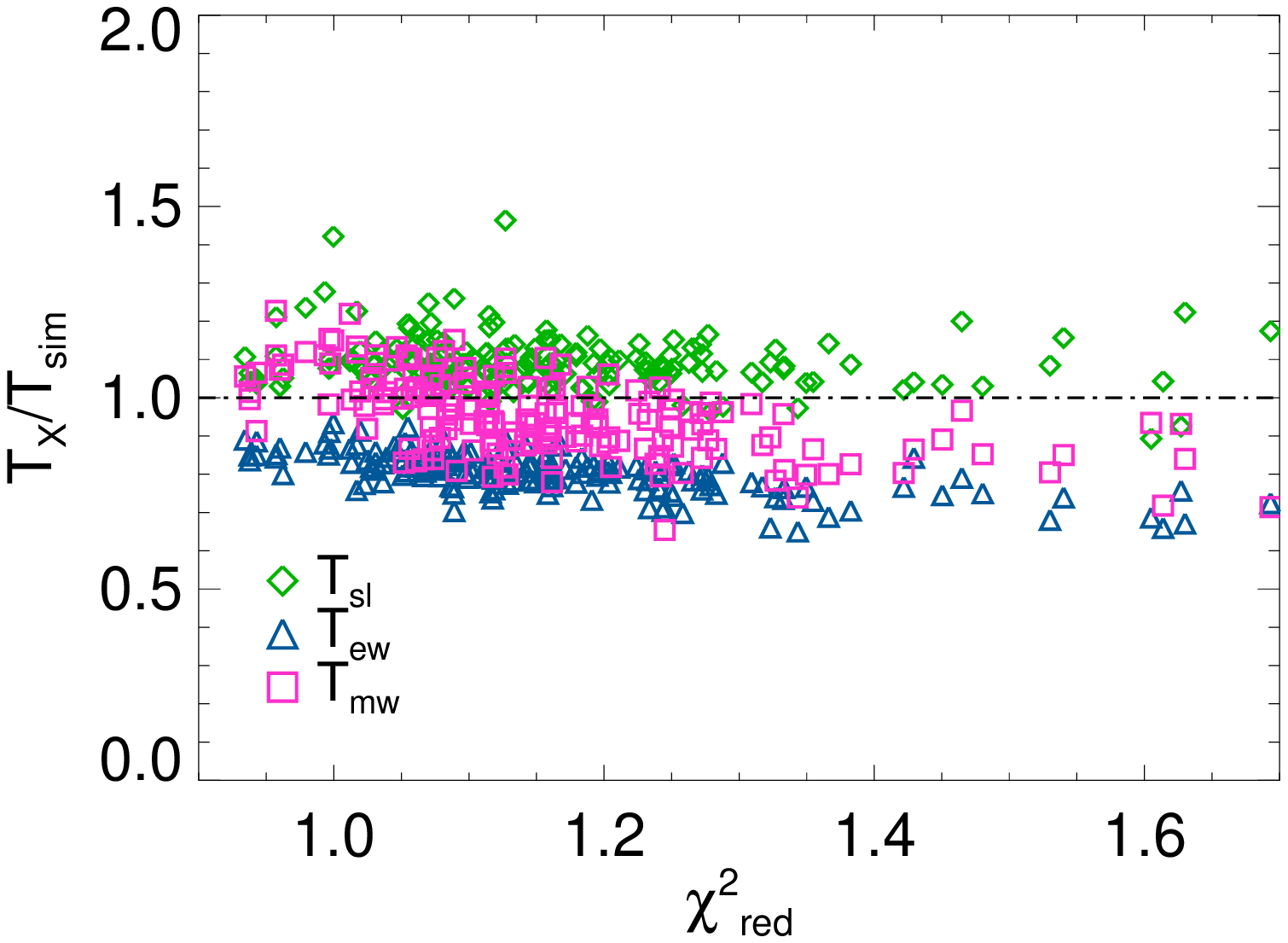}
\caption{Correlation of the temperature ratio $T_X/T_{sim}$ with the reduced $\chi^2$ obtained
for the single-temperature fit of the cluster Chandra spectra. The ratio is reported for the three theoretical estimates of temperature: $T_{sl}$ (green), $T_{ew}$ (blue), $T_{mw}$ (pink). The region considered is always that enclosed by $\rfive$ and the spectra are fitted over the entire $[0.5-10]\kev$ band.}
\label{fig:ratio_chired}
\end{figure}

We present here the results of the spectral fitting procedure (see \sec\ref{sec:xray_analysis}), 
in terms of the reduced chi-squared statistics, $\chi^2_{red}$.
\\
In \fig\ref{fig:chi2_fits} we show the $\chi^2_{red}$ distribution (as well as the cumulative one)
of the best-fitting single-temperature models for the cluster sample.
Additionally, we also show the dependence of the $T_X/T_{sim}$ ratio (see \sec\ref{sec:temp})
on the fit $\chi^2_{red}$.


\bsp

\label{lastpage}

\end{document}